\documentclass{aa}  
\usepackage{graphicx}
\usepackage{lscape}
\usepackage{txfonts}
\newcommand{\Gaia}{{\it Gaia}}
\newcommand{\usun}{$U_{\odot}$}
\newcommand{\vsun}{$V_{\odot}$}
\newcommand{\wsun}{$W_{\odot}$}
\newcommand{\kms}{km\ s$^{-1}$}

\begin{document}

\title{A self-consistent dynamical model of the Milky Way disc adjusted to {\it Gaia} data}
\author{A.C. Robin\inst{1}, O. Bienaym\'e \inst{2}, J.B. Salomon \inst{3,4,1}, C. Reyl\'e \inst{1}, N. Lagarde \inst{5}, F. Figueras\inst{6}, R. Mor\inst{6,7}, J. G. Fern\'andez-Trincado\inst{8}, J. Montillaud\inst{1}}
  \authorrunning{Robin et al.}
  \institute{
        Institut UTINAM - UMR 6213 - CNRS - University of Bourgogne Franche Comt\'e, France, OSU THETA, 41bis avenue de l’Observatoire, 25000, Besançon, France
        \and
        Observatoire Astronomique de Strasbourg, Universit\'e de Strasbourg, CNRS UMR 7550, 11 rue de l'universit\'e, 67000 Strasbourg, France
\and
Racah Institute of Physics, Hebrew University, Jerusalem 91904, Israel 
\and
Leibniz-Institut f\"ur Astrophysik Potsdam, An der Sternwarte 16, D-14482 Potsdam, Germany
\and
Laboratoire d'Astrophysique de Bordeaux, Univ. Bordeaux, CNRS, B18N, all\'ee Geoffroy Saint-Hilaire, 33615 Pessac, France 
\and
Dept. FQA, Institut de Ci\`encies del Cosmos, Universitat de Barcelona (IEEC-UB), Mart\'{i} i Franqu\`es 1, E-08028 Barcelona, Spain
\and
Pervasive Technologies s.l., c/Saragossa 118, 08006 Barcelona, Spain
\and
Instituto de Astronom\'ia, Universidad Cat\'olica del Norte, Av. Angamos 0610, Antofagasta, Chile
}
\date{}
\abstract
   {
 Accurate astrometry achieved by {\it Gaia}  for many stars in the Milky Way provides an opportunity to reanalyse the Galactic stellar populations from a large and homogeneous sample and to revisit the Galaxy gravitational potential.}
   {This paper shows how a self-consistent dynamical model can be obtained by fitting the gravitational potential of the Milky Way to
the stellar kinematics and densities from {\it Gaia} data.}
   {{We derived a gravitational potential using the Besancon Galaxy Model,}  and computed the disc stellar distribution functions based on three integrals of motion ($E$, $L_z$, $I_3$) to model stationary stellar discs. The gravitational potential and the stellar distribution functions are built self-consistently, and are then adjusted to be in agreement with the kinematics and the density distributions obtained from \Gaia\  observations. A  Markov chain Monte Carlo (MCMC) is used to fit the free parameters of the dynamical model to \Gaia\  parallax and proper motion distributions. 
   The fit is done on several sets of \Gaia\  data, mainly a subsample of the GCNS ({\it Gaia} catalogue of nearby stars to 100 pc) with $G<17$, together with 26 deep fields selected from eDR3, widely spread in longitudes and latitudes. }
   {We are able to determine the velocity dispersion ellipsoid and its tilt for subcomponents of different ages, both varying with $R$ and $z$. The density laws and their radial scale lengths for the thin and thick disc populations are also obtained self-consistently. This new model has some interesting characteristics that come naturally from the process, such as a flaring thin disc. The thick disc is found to present very distinctive characteristics from the old thin disc, both in density and kinematics. This lends significant support to the idea that thin and thick discs were formed in distinct scenarios, as the density and kinematics transition between them is found to be abrupt. The dark matter halo is shown to be nearly spherical. We also derive the solar motion with regards to the Local Standard of Rest (LSR), finding \usun=10.79 $\pm$ 0.56 \kms, \vsun=11.06 $\pm$ 0.94 \kms, and \wsun=7.66 $\pm$ 0.43 \kms, in close agreement with recent studies.}
   {The resulting fully self-consistent gravitational potential, still axisymmetric, is a good approximation of a smooth mass distribution in the Milky Way and can be used for further studies, including finding streams, substructures, and to compute orbits for real stars in our Galaxy.}

 \keywords{Galaxy:kinematics and dynamics, Galaxy:structure, Galaxy:evolution, Galaxy:disc, surveys
               }

\maketitle
\newcommand{\adisc}{$k,$}
\newcommand{\bdisc}{$\beta$}
\newcommand{\suytd}{$\tilde{\sigma}_{R_{ythd}}$}
\newcommand{\swytd}{$\tilde{\sigma}_{z_{ythd}}$}
\newcommand{\suotd}{$\tilde{\sigma}_{R_{othd}}$}
\newcommand{\swotd}{$\tilde{\sigma}_{z_{othd}}$}
\newcommand{\Rsud}{$\tilde{H}_{{\sigma_R}_{thin}}$}
\newcommand{\Rsuytd}{$\tilde{H}_{{\sigma_R}_{ythd}}$}
\newcommand{\Rsuotd}{$\tilde{H}_{{\sigma_R}_{othd}}$}
\newcommand{\Rswd}{$\tilde{H}_{{\sigma_z}_{thin}}$}
\newcommand{\Rswytd}{$\tilde{H}_{{\sigma_z}_{ythd}}$}
\newcommand{\Rswotd}{$\tilde{H}_{{\sigma_z}_{othd}}$}
\newcommand{\dvsd}{$\rho_{0_{thin}}$}
\newcommand{\dvsytd}{$\rho_{0_{ythd}}$}
\newcommand{\dvsotd}{$\rho_{0_{othd}}$}
\newcommand{\Rrhoratiod}{$\tilde{H}_{{\rho}_{thin}}$}
\newcommand{\Rrhoratioytd}{$\tilde{H}_{{\rho}_{ythd}}$}
\newcommand{\Rrhoratiootd}{$\tilde{H}_{{\rho}_{othd}}$}
\newcommand{\sigratio}{$\tilde{\sigma}_R/\tilde{\sigma_z}$}
\newcommand{\Lr}{$L_r$}
\newcommand{\Rsun}{$R_{0}$}
\newcommand{\Msun}{$M_{\odot}$}
\newcommand{\MH}{$[M/H]$}
\newcommand{\vtl}{$Vt_l$}
\newcommand{\vtb}{$Vt_b$}
\newcommand{\VTl}{$Vt_l$}
\newcommand{\VTb}{$Vt_b$}
\newcommand{\Rgal}{$R$}
\newcommand{\zgal}{$z$}
\newcommand{\sigR}{$\sigma_R$}
\newcommand{\sigz}{$\sigma_z$}
\newcommand{\GRP}{G$_{RP}$}
\newcommand{\MG}{M$_{G}^{*}$}
\renewcommand{\deg}{$^{\circ}$}
\newcommand{\Kepler}{{\it Kepler}}

\newcommand{\subf}[2]{%
  {\small\begin{tabular}[t]{@{}c@{}}
  #1\\#2
  \end{tabular}}%
}

\section{Introduction}

Understanding the Milky Way structure
via its dynamics is crucial in order to figure out Galaxy evolution, as the gravitation is the major force that drives Galaxy shaping. The mass distribution can be derived mostly from light for the baryons but completely relies on dynamical effects for the dark components. The construction of any realistic Galaxy model would need to simultaneously confront modelled visible matter, observed distributions and kinematics, and dark matter contributions to the kinematics via the gravitational potential. 

Therefore, self-consistent modelling approaches are needed that model the visible part, particularly the stellar populations, their imprint on the gravitational field, and how they feel the potential through their motions. The invisible components such as the dark halo have to be considered also with its visible effects on the rotation curve. The third major component, the interstellar matter, is only partly visible and constitutes an important uncertainty on the Galactic mass models, while its dynamics is notably different from the collisionless stellar dynamics. 

To understand how the Besan\c{c}on Galaxy Model (BGM) compares to the many existing modellings of the Galaxy, we emphasise that the term `Galactic dynamical model' covers several very distinct approaches. In general, published dynamical models try to find the mass distribution by means of the kinematics of stars,  or gas, clusters, satellite galaxies, stellar streams, and so on. These models propose a decomposition of the Galactic mass distribution into components: namely gas, stellar discs, halo,  and dark matter \citep[e.g.][]{1975PASJ...27..533M}, but they do not generally look for the dynamical consistency which involves each component.

A distinct approach to understanding the structure and history of the Galaxy is to look at stellar populations. For example, the TRILEGAL
 stellar population code \citep{2005A&A...436..895G} has been used to analyse the absolute colour--magnitude distribution using  stellar evolutionary tracks, allowing  determination of the Galactic star formation history \citep{2021MNRAS.506.5681D}. A complementary approach is the kinematical modelling (by opposition to  dynamical modelling) which consists in describing the stellar density and velocity distributions but without seeking dynamical consistency with the gravitational field. Thus, the Galaxia model  allowed the generation of a synthetic survey of the Milky Way \citep{Sharma2011ApJ...730....3S}.

In the context of our Galaxy, few models  exist that combine the stellar density and velocity distributions with the gravitational potential in a dynamically consistent way. We can mention   \cite{2011MNRAS.413.1889B}
 for the analysis of nearby stars and towards Galactic poles, and the studies of the stellar RAVE survey by  \cite{Sharma2014ApJ...793...51S},
\cite{2014MNRAS.445.3133P}, and \cite{2014A&A...571A..92B}.

Finally, we know of only three models of the Galaxy that combine these approaches and the implementation of dynamical consistency in a population synthesis model. These models describe the details of the stellar populations  with evolutionary tracks. These are the JJ model of \cite{2010MNRAS.402..461J} and \cite{2021A&A...647A..39S}, 
the ModGal model of \cite{ 2016MNRAS.461.2383P, 2018ApJ...860..120P}, 
and the Besan\c{c}on Galaxy Model, which is developed here.

\paragraph{Mass models and kinematics.}
Different methods have been used to investigate the overall mass distribution and dynamics of the Milky Way: from deriving density distributions and kinematics  independently, and computing forces from Jeans equations  \citep[][]{2021ApJ...916..112N}, 
to Schwarzschild modelling  \citep{2020ApJ...889...39V}, 
made-to-measure modelling  \citep{2015MNRAS.450.4050W},
and models based on  angle-action or integrals of motion  \citep{Binney2014MNRAS.439.1231B,2017A&A...605A...1R}. 

Contrary to \cite{2021A&A...646A..99R}, who investigate the deviations from axisymmetry  using \Gaia\ DR2 in order to investigate resonances and substructures, here we attempt to derive an axisymmetric model which reproduces the density and kinematics of the Milky Way in a wide range of Galactocentric distances. This model would represent the mean structure, i.e. the dominant pattern, and is also directly related to the overall potential and mass distribution. 

\cite{1998MNRAS.294..429D} developed a very popular axisymmetric mass model of the Milky Way based on a semi-analytical approach, but at this epoch, the data constraining the distribution functions were mainly limited to the measurement of the rotation curve and the kinematics of the local solar neighbourhood, with little information available on the radial motion of Milky Way satellites, leading to some unconstrained model parameters.
With the availability of \Gaia\  data \citep[DR1]{2016A&A...595A...2G}, \citep[DR2]{Brown2018A&A...616A...1G},\citep[eDR3]{2021A&A...649A...1G}, new developments of Galaxy models have arisen \citep{2017MNRAS.465...76M}. 
To mention just two from the literature, we refer the reader to  the new model by \cite{2022MNRAS.510.2242W} based on the motions of globular clusters from \Gaia\  eDR3, and the analytical model from the observations of cepheids in \Gaia\  DR2 data and other surveys \citep{2020ApJ...895L..12A}. 

Some studies also point towards non-axisymmetric structures coming from internal or external perturbations, such as the spiral feature detected by \cite{2018Natur.561..360A}, numerous stellar streams \citep{2021ApJ...914..123I,Malhan_2022}, and perturbations from the bar or spiral waves  detected in the solar neighbourhood from RAVE, APOGEE, or \Gaia\  DR2 \citep{2013MNRAS.436..101W,2015ApJ...800...83B,2018MNRAS.475.2679C,10.1093/mnras/stab3755,2021MNRAS.500.2645T}. However, the majority of the mass of the Galaxy is predominantly in smooth relaxed components, and this is why the development of axisymmetric self-consistent dynamical models of the Milky Way is useful.

Most axisymmetric models are based on analytical functions, such as the Miyamoto-Nagai formula \citep{1975PASJ...27..533M}, which was used by the popular model of 
\cite{1991RMxAA..22..255A}, 
the \cite{2016A&A...593A.108B} analytical model fitted to the gas rotation curve, or the \cite{2015ApJS..216...29B} model fitted to APOGEE data, among others.
The analytical functions are sometimes not flexible enough to follow the real forces at work in the Milky Way. 
\\
\paragraph{Dynamically consistent modelling.}
\cite{2014MNRAS.441.3284S,2016MNRAS.457.2107S} developed a method called `St\"ackel fudge' which allows the numerical determination of  approximate values of the  actions $ J$ of  stellar orbits (integrals of motion which have the dimension of angular momentum) in the case of an axisymmetric potential.  The method allows the distribution functions to be defined, and these are dependent on three integrals of motion (one of which is the angular momentum) for the stellar populations. This method, called $f({J})$ \citep{2020IAUS..353..101B}, has been used to model the densities and kinematics of various stellar samples from the RAVE survey by \cite{2014MNRAS.445.3133P}. In the present paper, we use a formally equivalent approach and define approximate integrals with the dimension of an energy \citep{Bienayme2015A&A...581A.123B,2019A&A...627A.123B}. These integrals allow us to build stellar disc distribution functions of densities and kinematics. This approach was  used for the analysis of stellar populations using  RAVE data \citep{2014A&A...571A..92B,2017A&A...605A...1R}, and the advantage it confers is that the integral expressions are  analytical, which presents a considerable simplification in terms of computation.

A self-consistent dynamical approach was used  by \cite{Bienayme1987a} who proposed solutions for the density and kinematics of the stellar populations valid only locally at the  position of the Sun. More recently,
\cite{Bienayme2015A&A...581A.123B} and \cite{2018A&A...620A.103B} developed a more general method to derive a dynamically self-consistent Galactic model assuming axisymmetry and using stellar distribution functions depending on integrals of motion. This method has also been extended to non-axisymmetric models \citep{2019A&A...627A.123B}.

In this paper, we attempt to derive a fully self-consistent dynamical model of the Galaxy based on stellar population synthesis modelling. We take up the method of \cite{2018A&A...620A.103B}  and apply it to the new \Gaia\  data release eDR3 ---where accurate astrometry is available--- to characterise the kinematics of stellar populations in a large volume. This approach allows us to confront the dynamics with observations of the stellar motions and their spatial distributions, and therefore to test the full 6D-space distribution functions. 
The paper is set out as follows. In Sect.~\ref{Sect:Dynamics} we present the principle of the model and in Sect.\ref{Sect:fit} we explain  how self-consistency is obtained. In Sect.~\ref{Sect:data} we show the data selection from the \Gaia\  eDR3 while in Sect.~\ref{Sect:simu} we describe the simulations and the MCMC strategy  used to derive the model parameters. Results are presented in Sect.\ \ref{Sect:result}. The new dynamical model characteristics are presented in Sect.\ \ref{Sect:newmodel}. In Sect.~\ref{Sect:discussion} we discuss these results in light of previous studies, both theoretical and observational. We outline our conclusions and perspectives in Sect.\ \ref{Sect:conclusion}.

\section{Dynamics in the BGM}
\label{Sect:Dynamics}

The BGM  follows a population synthesis approach. It assumes that the Galaxy is made of several stellar components, mainly a thin disc, a thick disc, a bar, and a stellar halo, to which non-stellar components are added: an interstellar matter disc, a dark matter halo, and a central bulge. The population synthesis allows us to compute catalogue simulations for the stellar components which are based on assumptions describing the star formation (initial mass function (IMF), star formation history (SFH)), and evolution (evolutionary tracks). For generated stars in the simulations, observables are computed using atmosphere models, while a 3D extinction map is used to account for absorption and reddening for every simulated star, and error models are added on observables. A dynamical model is then used to compute star kinematics in a self-consistent manner, that is the total mass distribution of all the components (stellar and non-stellar) is used to compute the gravitational potential of the Galaxy, which in return consistently provides the distribution functions used in computing the observables of the stellar components. 

\subsection{New dynamical modelling}

Previous versions \citep{Bienayme1987a,Robin2003} of the BGM proposed a restricted dynamical consistency. In these earlier versions, the mass distribution of all Galactic components was used to reproduce the Galactic rotation curve. The vertical density distributions of stellar discs  was constrained by their vertical velocity dispersions and the vertical variation of the gravitational potential through the Jeans equation. However, this constraint linking the thickness of the stellar discs to the vertical velocity dispersion was only applied at the solar Galactic radius \Rsun. These density laws were mainly \cite{1979IAUS...84..451E} laws
which have very similar shapes to dynamically consistent density laws, which are laws  that depend on three integrals of motion
\citep[Fig. 1-2][]{2018A&A...620A.103B}. 

In the new BGM version presented here, the dynamical consistency is not restricted to the Galactic radius \Rsun\ but is obtained with remarkable accuracy at all Galactic radii \Rgal\ larger than 4\,kpc, and for distances up to 6 kpc away from the Galactic plane. 

Moreover, the density and kinematic laws of each stellar disc are no longer modelled by empirical laws, but  expressed with a function of three integrals of motion ($E$, $L_z$ and $I_3$). This  leads to an exact stationary representation of the position and velocity distributions of the stellar thin and thick discs. 
Our $I_3$ integral is partly  analogous to the integrals of  St\"{a}ckel potentials, with a similar approach to the work by \cite{2014MNRAS.441.3284S}  but here analytically calculated 
\citep{Bienayme2015A&A...581A.123B}.

To this end, we adopt observational constraints for the rotation curve (Sect. \ref{Sect:rotation}), and define mass components, some of them being fixed (Sect. \ref{Sect:mass-comp}) and others adjusted in the process (Sect. \ref{Sect:free-comp}). 

\subsection{Adopted Galactic rotation curve}
\label{Sect:rotation}

Our adopted rotation curve is built piece by piece.  Below $R$=2 kpc, we do not fit the rotation curve because the gas motions are dominated by non-axisymmetric motions.
Between 2 and 6 kpc, we use  the \cite{2018RNAAS...2..156M}  HI velocity curve based on recent data and the up-to-date values of the Sun-Galactic centre distance $R_0$ = 8.122 kpc and the Local Standard of Rest (LSR) velocity at the Sun $V_0$ = 233 \kms. From 6 to 10 kpc, we use the recent determinations of the velocity curve based on Cepheids and DR2 data \citep{2019ApJ...870L..10M, 2020ApJ...895L..12A}. We set $V_0$ =233\kms and $\nabla V_c$ = -1.35 \kms kpc$^{-1}$.
For the outer Galaxy, that is, $R$>10 kpc up to 60 kpc, we consider a decreasing rotation curve proposed by \cite{2020MNRAS.494.4291C}  
\citep[see Fig. 6 of][]{2021MNRAS.501.5964D}. This implies that our mass model has nearly the same Galactic mass at large radii as the model of \cite{2020MNRAS.494.4291C} $M (<100$ kpc) = 6.1 $\times 10^{11}$ \Msun.
The distance of the Sun from the Galactic centre is taken to be $R_0=8.1$ kpc, a rounded value in agreement with the determinations of 
\citet[][$R_0=8.122$ kpc]{2018A&A...615L..15G}, 
\citet[][$R_0=8.178$ kpc]{2019A&A...625L..10G}
and
\citet[][$R_0=8.1$ kpc]{2021AstL...47..607B}.

\subsection{Mass components of the Galactic model}
The  Galactic mass ingredients are the baryonic components,  stars, interstellar matter (ISM), and a dark matter halo. The mass of these components allows us to compute the total Galactic gravitational potential and we use this potential to build dynamically consistent density and kinematics for each stellar disc component. 
The observed and fitted rotation curves are shown in Fig. \ref{fig:vcirc}, together with the contribution of the baryonic and dark matter components.

\subsubsection{Fixed components}
\label{Sect:mass-comp}
The shape of a few components remains fixed in the present work: the stellar halo, the ISM, and the bar, with characteristics already defined  \citep{Robin2003,Robin2012a}.

The ISM density distribution is a double exponential law:
\begin{equation}
\label{eq:ISM}
\begin{tabular}{ll}
$\rho_{\mathrm{ISM}}$ &= $\rho_c \exp(-R/h_R) \exp(-|z|/h_z)$
\\
with &  $h_R$=7000 pc and $h_z$=200 pc. \\
\end{tabular}
\end{equation}

Its local density is $\rho(R_0)=0.0275 M_\odot\,pc^{-3}$, and the local surface mass density $\Sigma_{\mathrm{ISM}}(R_0)= 11 M_\odot \, pc^{-2}$. The mass density of the stellar halo is given by:
\begin{equation}
\label{eq:stellarhalo}
\begin{tabular}{ll}
$ \rho(R,z)= C_1 \, a^{-2.44}$ &if $a>500$ pc,\\
$ \rho(R,z)= C_1 \, 500^{-2.44}$&if $a<500$ pc,
\end{tabular}
\end{equation}
with $a=  \sqrt{R^2+z^2/q_{\mathrm{stellar}}^2}$, $q_{\mathrm{stellar}}=0.76$, and the local density $\rho(R_0,0)=9.32\times10^{-6} M_\odot \ pc^{-3}$ is used to fix the constant $C_1$.

\begin{figure}
\begin{center}
\includegraphics[width=9cm]{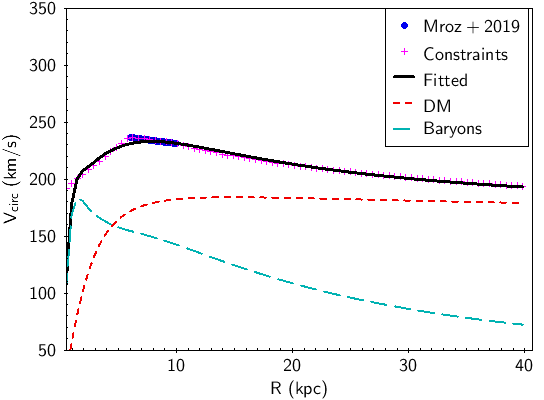}
\caption{Adopted Galactic rotation curve. Blue dots are cepheid observations from \cite{2019ApJ...870L..10M}. Magenta crosses represent the composite curve used as a constraint to the fit (see text). The full black line represents the circular velocity of our best-fit model. The dark matter contribution is shown as a short-dashed red line, and the baryon contribution, including the central mass, as a long-dash cyan line.}  

\label{fig:vcirc}
\end{center}
\end{figure}

We add a central bulge ---whose mass density is given by a
Plummer sphere--- in order to help adjust the rotation curve. This component can be partly stellar and partly dark halo. In Fig. \ref{fig:vcirc}, it is included in the baryonic component:

\begin{equation}
\label{eq:bulge}
\rho(R,z)= \frac{C_2}{ (R_{\mathrm{bulge}}^2 + R^2 +z^2)^{2.5}}
,\end{equation}
with $R_{\mathrm{bulge}}$=1 kpc and $C_2$ a normalisation constant.
\\
The total mass of these components is $20\times 10^9$ \Msun\  for the bulge,   10.8$\times 10^9 $\Msun\ for the ISM, and  317 $\times 10^6$ \Msun\  for the stellar halo inside 20 kpc and 318 $\times 10^6 $\Msun\ inside 100 kpc.
\\

\subsubsection{Dark halo}
\label{Sect:free-comp}

To model the Galactic rotation curve, we include a dark matter spheroid.
The free parameters of this  dark halo are the core radius $R_{\mathrm{DM}}$, the central density $\rho_c$, and the spheroidal flattening $q_{\mathrm{DM}}$. We point out that, with a  flattening $q_{\mathrm{DM}}$ not too different from 1, the density of our halo can be negative close to the $z$-axis but sufficiently far from the Galactic plane so that it remains realistic in our domain of interest.

To model a decreasing rotation curve 
 at large $R$  we use a  dark matter potential given by:
\begin{equation}
\label{eq:DM_dec}
\begin{tabular}{llll}
$\Phi_{\mathrm{DM}}$ &=$ - 4 \pi G \rho_c K \left( R_{\mathrm{DM}}^2+R^2+z^2/q_{\mathrm{DM}}^2  \right)^{-\gamma} $
\\
with &$K=\frac{R_{\mathrm{DM}}^{2+2 \gamma}}  {2\gamma (2+1/q_{\mathrm{DM}}^2)}$
\end{tabular}
.\end{equation}

We fixed the exponent $\gamma=0.05$ of this potential to be able to adjust the decreasing rotation curve for the large values of $R$ from 20 to 50 kpc
\cite[][]{2021MNRAS.501.5964D}. 
In Table~\ref{tab:DM}, the values of the parameters for the dark matter halo are presented.

\subsection{Stellar disc distribution functions} 

The stellar discs are modelled using the population synthesis scheme \citep{Robin2003} with a thin disc made of seven subcomponents of different ages and a thick disc made of two components, young and old \citep{2017A&A...605A...1R}.
For each stellar disc component, the free parameters to determine are the solar position values for the density, the vertical velocity dispersion as a function of age, and the radial scale lengths. 
The complete stellar disc distribution functions are constrained by the dynamical consistency (see below) and by the fit to the stellar counts and kinematics.

Our stellar population distribution functions (DFs) are 3D generalisations of the \cite{1969ApJ...158..505S} distribution functions for an axisymmetric gravitational potential.
The DFs of positions and velocities of each stellar disc are modelled with a function of integrals of motion and are written as follows:

\begin{equation}
\label{eq:DF}
    \begin{tabular}{llll}
$f(x,v) \sim g(L_z)  \, \tilde\rho_{0} \, \exp\left({\frac{R_c-R_0} {\tilde{H}_{\rho}}}\right) \,  \exp\left({-\frac{E_{\parallel}}{\tilde{\sigma}_{R}^2}}\right)  \,  \exp\left({-\frac{E_{\perp}}{\tilde{\sigma}_{z}^2}}\right) $
 \end{tabular}
 ,\end{equation}
 
\begin{equation*}
    \begin{tabular}{llll}
with &
  $\tilde{\sigma}_R = \tilde{\sigma}_{0,R}  \exp\left({-\frac{R_c-R_0}{\tilde{H}_{\sigma_R}}}\right)$\\
and &
    $\tilde{\sigma}_z = \tilde{\sigma}_{0,z}  \exp\left({-\frac{R_c-R_0}{\tilde{H}_{\sigma_z}}}\right)$, \\
\end{tabular}
\end{equation*}

where  $R_0$ is the Galactic radius at solar position,  $R_c(L_z)$ is the radius of the circular orbit with angular momentum $L_z$, and $E_{\parallel}\sim E(1-I_3)$ and $E_{\perp}\sim I_3$ are integrals of motion depending on the energy $E$ and a third integral $I_3$ 
\citep[see][]{2018A&A...620A.103B}, which are linked respectively to   the radial and vertical motion of the stars.

The distribution function $f(x,v)$ allows us to compute its various moments, which give us the density $\rho(R,z)$, the rotational velocity $V_{\phi}$, the velocity dispersions $\sigma_R$, $\sigma_\phi$, and
 $\sigma_z$, and the tilt angle of the velocity ellipsoid. All the quantities that depend on position $R$, $z$ , and disc components are stored in a table. When a simulated catalogue of stars is created, the kinematical properties of each star are drawn according to the characteristics given in this table.
 
For each stellar disc, the input parameters of the distribution function are noted with a tilde in Eq.~\ref{eq:DF}. These are: $\tilde\rho_{i,0}$, $\tilde\sigma_{i,R_{0}}$, $\tilde\sigma_{i,z_{0}}$,  $\tilde{H}_{i,\rho}$,  $\tilde{H}_{i,\sigma_{R}}$, and  $\tilde{H}_{i,\sigma_{z}}$, with the index {\it i} of the disc component. 

The three first input parameters  are  directly related  to the computed moments of the DF, the computed density and dispersions at the solar position, respectively  $\rho(R_0,z=0)$, $\sigma_R(R_0,z=0)$, and $\sigma_z(R_0,z=0),$ but they do not have exactly the same values.

The other free parameters, $\tilde{H}_{\rho}$, $\tilde{H}_{\sigma_R}$, and $\tilde{H}_{\sigma_z}$, are related to the scale lengths for the density and for the kinematics. The exact scale length can be computed from the tabulated DF. The density and kinematical laws have a nearly radial exponential decrease beyond $R=4$ kpc and the computed scale lengths vary with $z$.

At large radii, we use DFs that are slightly different from Eq~\ref{eq:DF}. A threshold of 5 \kms\, is imposed for the velocity dispersions at very large $R$ where we do not expect that the velocity dispersion becomes smaller than that of the ISM. On the other hand, for radii $R$ smaller than 4 kpc, the velocity dispersions are set to an almost constant value to avoid overly large values. These modifications to make the DF more realistic have no impact on the results presented here because they cover domains where we do not make comparisons with \Gaia\  data.

The other dynamical constraints reside in reproducing the Galactic rotation curve for $R >$ 4 kpc and the constraint on the local density of dark matter  $\rho_{dm}(R_0,z=0)=0.010 M_\odot \, pc^{-3}$
in the solar neighbourhood 
\citep{2014A&A...571A..92B,2020A&A...643A..75S}. This  constraint is mainly satisfied  by modifying the flattening of the dark matter halo.
The consistency of the stellar distribution functions  and force fields is achieved with an accuracy of the order of one per thousand
\citep[see Fig. 1-2 in][]{2018A&A...620A.103B}.

We emphasize that with this new version of the BGM, the number of free parameters of the model is reduced, and the asymmetric drift, the azimuthal velocity dispersions $\sigma_\phi$, and the tilt angle of the velocity ellipsoid are fully constrained by the dynamical consistency (now, the tilt of the velocity ellipsoid depends not only on the position but also on the stellar population). 

\section{Fitting process}
\label{Sect:fit}
\begin{figure*}[h]
\begin{center}
\includegraphics[width=1\textwidth]{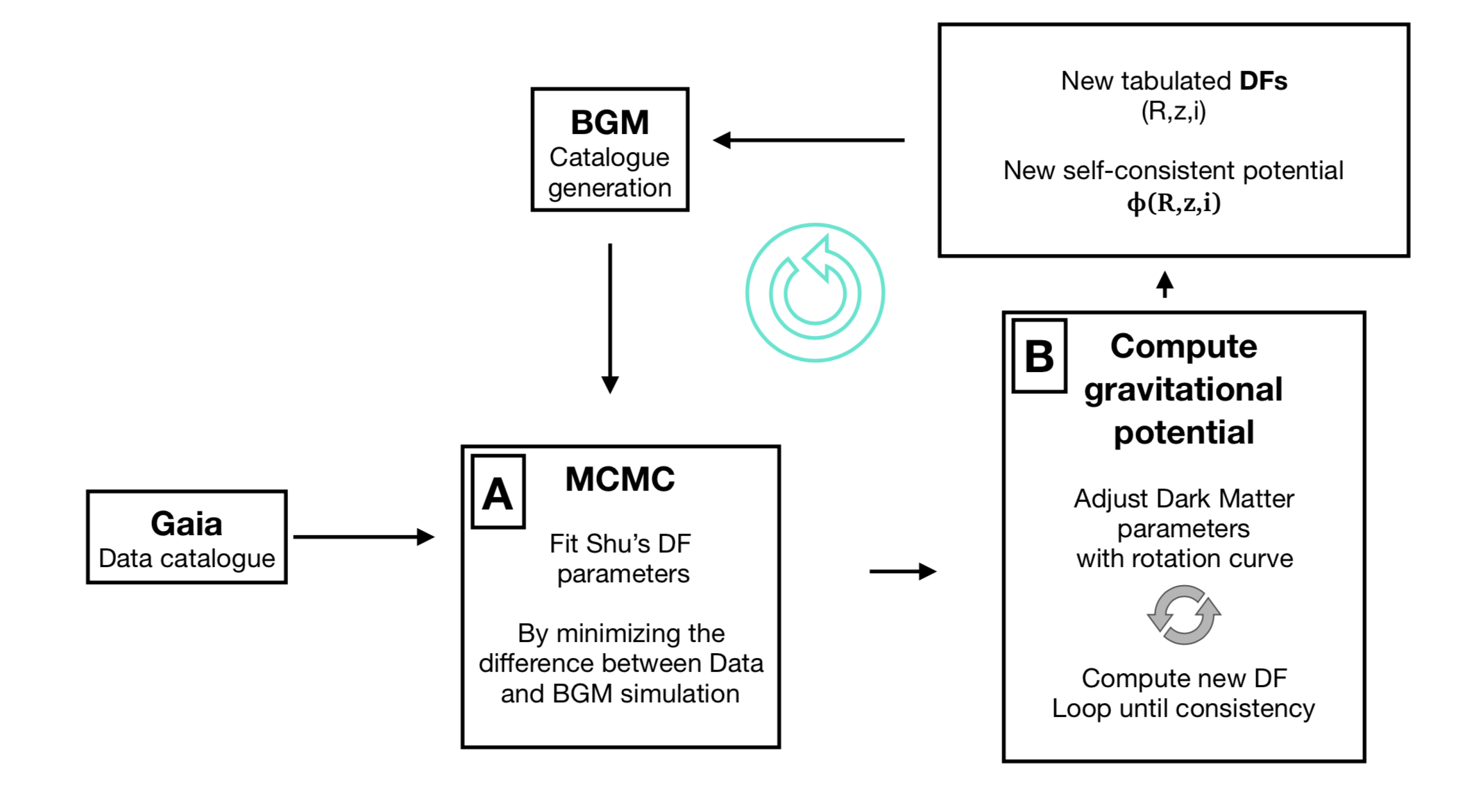}
\caption{Scheme of the fitting process. }
\label{fig:loops}
\end{center}
\end{figure*}

The scheme of the fitting process is summarised in Fig.~\ref{fig:loops}.

With a given set of model parameters as given in Sect. \ref{Sect:Dynamics}, we compute an exact self-consistent dynamical model and stationary disc DFs. 
We determine the mass distribution of the Galactic model by summing up the density of all components of the model: stellar discs, stellar halo, bulge, ISM, and dark matter halo. The potential is obtained by solving the Poisson equation with the boundary conditions given by  a direct calculation of the potential on a rectangular contour located at 60 kpc in $R$ and 10 kpc in $z$. The parameters of the dark matter halo are adjusted so that the Galactic model reproduces the Galactic rotation curve.

We then process the fitting to the Gaia data in two steps:
\begin{itemize}
\item Step A: 
With a Markov Chain Monte Carlo we modify the Shu's parameters to fit the observed kinematics and density distributions by comparing the simulated catalogue to Gaia data (see Sect. \ref{Sect:simu}). This fit is achieved with a simplified, fast, but approximate version of the dynamical model.
\item Step B: The parameters obtained with the previous best fit are used to compute a new exactly self-consistent potential and  stationary DFs.
\end{itemize}

After step B, we loop on Step A with the revised potential and we follow the improvement of the likelihood with regards to a previous self-consistent model. 
We iterate steps A and B until there is no improvement of the global likelihood. We monitor the values of scale lengths and verify that densities have not significantly changed (i.e. are within 1 sigma) compared with the exact self-consistent DFs (to ensure that the approximate formulae used are valid).

The exact calculations of the potential and DFs take about ten CPU minutes on a single processor, an amount of time that does not allow the dynamical consistency to be recalculated at each of the tens of thousands of steps in the MCMC chains. To circumvent this difficulty, we developed an approximate and fast version of the dynamical model calculation. This latter allows us to build alternative approximate computations of the DFs from an already consistent dynamical model, provided that the Shu's parameters are only slightly modified, that is, by about 20 per cent at most. This is a partially linearised version of the DF computation of a consistent dynamical model in the immediate vicinity of another one. The advantage of this approximation of the calculation is that it takes the CPU less than a millisecond instead of a few minutes for an exact computation.

For the MCMC chain (step A), the simplified version of the DF computation is used and when the Shu's parameters are modified by more than 20\%, the exact calculation of the dynamical consistency is repeated (step B). This way of proceeding allows us to converge more quickly towards the maximum likelihood and to determine with precision the confidence interval of the various free parameters.

\section{Data selection}
\label{Sect:data}

The comparison of sky densities and kinematics between models and data requires that the data completeness be ensured or that the selection functions of the data be accurately determined and reproduced in the model simulations. For the purpose of constraining  the distribution functions of the Milky Way, we used \Gaia\  eDR3 data \citep{2021A&A...649A...1G} and selected stars in the range of apparent $G$ magnitude between 6 and 17. These magnitude limits are determined to ensure that most of the stars have very reliable parallaxes and proper motions and the sample is as complete as possible. We have not considered radial velocities  here because in eDR3 they concern stars brighter than about 12, restricting the data sample too much. Instead, we considered the proper motions and parallaxes and computed the projected tangential velocities  along Galactic longitudes and latitudes \vtl\ and \vtb\ defined in Equations~\ref{eq:VT} :
\begin{equation}\label{eq:VT}
      Vt_l =\mu_{l*}\times 4.74/\varpi\\
        Vt_b =\mu_{b}\times 4.74/\varpi,  
\end{equation}
where $\mu_{l*}$ refers to $\mu_{l}\times cos(b)$. $\varpi$ is the parallax in milliarcseconds and $\mu_l$  and $\mu_b$ are the proper motions in Galactic coordinates in milliarcseconds per year. With the value of the constant used (4.74), the transverse velocities are in \kms.

We make use of two different Gaia samples: a local sample, and a selection of deep fields.
In order to best determine the Solar motion and the local densities of various populations, we used the \Gaia\  Catalogue of Nearby Stars (GCNS, \citep{2021A&A...649A...6G}), which has a very well-defined selection function, has accurate proper motions and parallaxes, and has a completeness of over 99\% for $G<19$ \citep{2021A&A...649A...6G}. The astrometric accuracies in our selected sample (with $G<17$) are (0.029, 0.026) mas\ yr$^{-1}$ along the two proper motion axes, and 0.028 mas on parallaxes. It also depends on the position on the sky due to the scanning law. This is described on the \Gaia\  website\footnote{\url{ https://www.cosmos.esa.int/web/gaia/science-performance}}. On average, the achieved tangential velocity accuracy is about 0.015 \kms. 

This whole sky local sample containing stars up to 100 pc is then divided into 40 subfields: six bins in latitude with steps of 30\deg\ and eight bins in longitude with steps of 45\deg (for |b|<60\deg) ---or four bins in longitude at the poles for a better statistics--- in order to be able to measure tangential velocity distributions in different directions.

We note that we are not assuming that this sample is an homogeneous sphere. We instead simulate the sample selecting parallaxes larger than 10 mas after applying observational errors, accounting for the different scale heights of different populations, which imply variations in density depending on age. However, we cannot account for density fluctuations due to spiral arms or other substructures that could be present in the GCNS, because our model is axisymmetric.
\\
To constrain the distribution function outside the local sphere, we also selected \Gaia\  data in cardinal directions (longitudes 0\deg, 90\deg, 180\deg and 270\deg) and various latitudes (0\deg, $\pm$ 20\deg, $\pm$ 45\deg, $\pm$ 60\deg\ and the poles). The data were selected from the \Gaia\  archive within a given radius around each field centre (radius of 1.414 \degr\ for latitudes $|b|\leq 20$\deg, 3\deg\ for latitudes $\pm$45\deg, 5\deg\ for $b=\pm$60\degr and 8\degr\ at the poles to ensure a reliable statistics. We disregarded the Galactic centre field where data are noticeably incomplete due to crowding. 

Looking carefully at fields where simulated colour distribution disagreed with the data in the space (colour vs. parallax), we identified the parallax  at which there is a cloud of extinction that is not well modelled. In this way, we distinguished two fields (l=270\deg, b=0\deg, and l=180\deg, b=0\deg) where (in the data) a cloud leads to changes in the median colours at parallaxes <0.7mas. For these fields we limited the comparison to parallax>0.7mas.
We identified a field with a large discrepancy between model and data which may be due to the Monoceros Ring \citep{2003MNRAS.340L..21I,GCS} or to the warp, as a matter of debate: when comparing parallax distributions between l=180\deg, b=20\deg\ and l=180\deg, b=$-20$\deg, an excess of stars can be seen clearly in the north field at parallaxes $<$ 0.7 mas with respect to the south field. 
Therefore, we also disregarded stars in this field in the global comparison. We also eliminated the field at l=90\deg, b=0\deg, where our extinction model (see Sect.\ref{Sect:simu}) is insufficient to explain the CMD.
We finally have 39 subfields in the local sample and 26 deep fields.

In order to help obtain information on the ages of the stellar population, we also make use of a pseudo-absolute magnitude, defined as:
\begin{equation}\label{eq:MG}
      M_G^* = G + 5 \times \log_{10}(\varpi \times 1000.) +5,
\end{equation}

where the parallax $\varpi$ is in units of mas, and \MG\ corresponds to the true absolute magnitude when extinction is negligible and the parallax error is small. Because the simulations are done in observable space, including extinction and observational errors, the simulated \MG\ is directly comparable with the observed one. We selected stars in the range 1 to 7 mag in \MG\ in order to avoid low-mass stars (masses below $\approx$ 0.6 \Msun), which are not well simulated in the BGM, due to a lack of good stellar models for low-mass stars. 

For each field, we then used the pseudo absolute magnitude \MG\ between 1 and 7 divided into three bins, and further selected $\varpi>0.4 $ mas to avoid distant regions where the parallax contains very little information on the distance, apart from in the three fields with Galactic coordinates (270\deg,0\deg), (180\deg,0\deg), and (180\deg,20\deg) for which stars are selected with $\varpi>0.7 $ mas (see above). Our sample contains a total of 545\,280 stars, that is 44\,580 in local fields extracted from the GCNS and 500\,700 stars in the deep fields. 

\section{Simulations and MCMC strategy}
\label{Sect:simu}

\subsection{Basic parameters}
Simulations of the data samples are done using a revised version of the BGM, where evolutionary tracks have been updated using STAREVOL library \citep{2017A&A...601A..27L,2019A&A...621A..24L} 
for stellar masses larger than 0.6 \Msun,
  while the IMF and SFH of the thin disc were determined from an analysis and fit to \Gaia\  DR2  \citep{2018A&A...620A..79M,2019A&A...624L...1M}. 
The version of the BGM used in these works is referred to as Mev2011. 
\\

In the BGM scheme, the stars are generated from a mass reservoir from which their mass is withdrawn. The quantity of mass in the mass reservoir is computed according to the stellar density of the subcomponent (the seven age bins of the thin disc and two age bins for the thick disc) in the volume element considered (defined by $R$ and $z$ and the geometry of the cone for the direction of observation).

The  mass of each star is drawn following a three-slope IMF, and its age is drawn in the age range considered for the subcomponent considered. The metallicity is also drawn according to the assumed age--metallicity relation. Then the star is followed on the corresponding evolutionary track interpolated in mass, age, and metallicity in the grid and, if the age is not greater than the theoretical stellar lifetime, the astrophysical quantities of the star (temperature, luminosity, gravity, radius) are obtained from the corresponding interpolated tracks.

In previous model versions (from \cite{Czekaj2014}), the mass in remnants was estimated from the SFH and was subtracted before the start of the star-generation process. In the present version, for the sake of consistency, the stars that are found to be at the end of their life are treated specifically: we compute both the mass of gas released into the ISM and the mass subsisting in the remnant using the initial-to-final-mass ratio from \cite{2009MNRAS.395.1857K}. 
The mass regained by the ISM is added to the mass reservoir of the same age subcomponent, assuming instantaneous recycling. In this version of the model, we do not follow the stars over theoretical white dwarf (WD) tracks; they are instead generated as in previous BGM versions using pre-computed Hess diagrams  but with updated WD luminosity functions \citep{2005ApJS..156...47L}.

In simulations, we take into account binaries by drawing stars in the mass of gas available at a given position; first singles and primary stars, and then secondaries with a proportion which depends on the primary mass, as explained in \cite{Czekaj2014}. The binary fraction and distributions of semi-major axis and eccentricities follow the prescription of \cite{2011AIPC.1346..107A}. After the simulation is done and the apparent separation of the binary components is computed, we assume that, as in \Gaia\  data, the binaries are separated when their projected distance is larger than 0.4 mas. Otherwise, we merge the two components and attribute the total flux
in each photometric band to the unresolved system. The kinematics of the binary system is the same as the two components, neglecting orbital effects.

In this self-consistent version, as explained in Sect.\ref{Sect:Dynamics}, the density laws in the discs are self-consistently computed from stellar DFs and the gravitational potential. They are available as tables which are interpolated during the simulation. The kinematics of each star is also obtained from tables at any position in (\Rgal, \zgal) and for each subcomponent of the disc. As in previous versions, the thin disc has seven subcomponents of different ages between 0 and 10 Gyr, and the thick disc is modelled as two components: the so-called young thick disc for which the SFH follows a truncated Gaussian centred on 10 Gyr, with a standard deviation of 2 Gyr truncated between 8 and 12 Gyr, and the old thick disc with a SFH centred on 11 Gyr, with a standard deviation of 1 Gyr and truncated between 10 and 13 Gyr \citep{nasel2018}. The initial kinematical parameters and density laws for the thick discs were taken from the fit to RAVE survey data combined to \Gaia\  DR1 \citep{2017A&A...605A...1R}.

In simulations, we make use of a 3D extinction map, which is a combination of the whole sky map of  \cite{2019A&A...625A.135L}, called the {\it Stilism} map, and the 3D map provided by \cite{Marshall2006} which covers latitudes $|b|<$+10\deg  but extends further to about 10 kpc.
For the purpose of continuity between the two maps, the {\it Stilism} map has been modified to use Marshall's map as a prior for running a specific solution from the inverse method used to build {\it Stilism} (Lallement, private communication). This  specific solution  is used in our simulations.

\subsection{Simulations of \Gaia\  data samples}

Proper motion and parallax errors are simulated as random errors following Gaussian distributions assuming an error determined by the \Gaia\  DPAC, the values of which depend on magnitude, colour, and position on the sky (see Sect.\ref{Sect:data}). These random errors are added on the simulated motions in order to best reproduce the data.

We apply the same selection for \MG, colours, and parallaxes to the simulations and the data. Those initial simulations are referred to here as `mother simulations' and are modified during the fitting process. The simulated catalogues are completely recomputed after each Step B, as explained in Figure \ref{fig:loops}, while they are only modified (kinematics of each star recomputed and weights applied to each star according to the new density law) during the MCMC fitting process (step A).

\subsection{MCMC strategy}

Data and simulations are divided on the sky in bins corresponding to directions (39 in local sample, 26 in deep fields), and along each direction in bins of logarithm of the parallaxes. For each of these subsamples, we compute the quantiles of the distributions in \vtl\ and \vtb\ (0.1, 0.25, 0.5, 0.75 and 0.9 quantiles), and compare their values between model and data. The goodness of fit is estimated from the sum of the square of the difference between model and data divided by the standard deviation, for each quantile and for each of the transverse velocity coordinates in each bin. In this process, which relates to approximate Bayesian computation \citep[ABC-MCMC]{marin:pudlo:robert:ryder:2011}, the likelihood is not computed analytically, as it would be too complex.

The $log(\varpi)$ distributions for parallaxes larger than 0.4 mas (0.7 in the case of three specific fields; see above) are binned in steps of 0.1, giving 15 bins in deep fields. The relative difference of the $log(\varpi)$ distribution between model and data is added to the goodness of fit with a normalisation factor so that both constraints (goodness of fit of the quantiles of velocity distributions and $log(\varpi)$ relative difference) are roughly of the same order in most fields. In this way, the density distributions and the kinematics are constrained simultaneously.
\\
The density parameters considered in the MCMC fit are the local density in the thin and thick discs, and the Shu's parameters as explained in Sect.~\ref{Sect:fit}. For kinematics, we also fit the three components of the Solar motion \usun, \vsun, and \wsun.
To avoid an excessive number of free parameters and degeneracies, we considered that the Shu's scale lengths for the thin disc are globally modified by a single factor for all age components, as well as the radial to vertical velocity dispersion ratio \sigratio. Moreover, instead of fitting a \sigR\ for each age component, we assume that the age--velocity dispersion relation (AVR) follows a power law with two free parameters:

\begin{equation}\label{eq:avr}
\tilde{\sigma}_z (\tau) = k \times \tau ^\beta
,\end{equation}
 where $\tau$ is the age of the population in gigayears, and $k$ and $\beta$ are the fitted parameters. 
For each age subcomponent, we take the mean age for the value of $\tau$.

{We impose that the relative distribution of local stellar densities as a function of the age is given by the SFH discussed in Sect.~\ref{Sect:simu}. The local thick-disc density remains a free parameter determined during the fitting process. Old thick-disc parameters were first considered to be fitted but were not sufficiently constrained with the set of data used here, being minor everywhere. }

The fitting process is done in several steps as already explained in Sect.~\ref{Sect:Dynamics} and summarised in Fig. \ref{fig:loops}.  Below, we present the results obtained after several iterations on steps A and B until convergence.

\section{Results}
\label{Sect:result}

Table~\ref{tab:res-B7jul} presents the parameters fitted by the MCMC process during step A of the last loop; the range of parameter values (min and max) that we allowed in the chains; and the median and uncertainties determined using the tail of the best Markov chains.  
 The uncertainty is computed as the difference between the third and first quartiles. 

Thanks to the large sample of local stars, we were able to determine the solar motion with very good accuracy, and the same is true for the velocity dispersions for the thin and young thick discs (within 2 \kms), as seen in Table~\ref{tab:res-B7jul}.
For Shu's parameters and densities, we derive parameter values relative to the  exact self-consistent model obtained in Step B. The parameter range is therefore 0.8 to 1.2, corresponding to the 20\% range where the simplified computation of the new model is accurate enough (Sect.~\ref{Sect:Dynamics}).
We see that the values are all in this range and compatible with the value of 1 within 1 sigma. Therefore, these models are the final models and cannot be improved further with this data set. As seen in Table~\ref{tab:res-B7jul}, a reasonable accuracy of 10\% to 15\% is reached on the density and velocity dispersion scale lengths and local density.

\begin{table*}[h!]
\caption{Parameters determined by MCMC. The third and fourth columns indicate the minimum and maximum values imposed during the last runs of the MCMC (see text). The median and uncertainty are given for each fitted parameter. The uncertainties are computed from the difference between the third and first quartiles. Parameters below the line are correction factors of the Shu's parameters from the last self-consistent dynamical model (see text). 
}
\label{tab:res-B7jul}
\begin{center}
\begin{tabular}{llllllll}
\hline
Parameter &  Unit & Min & Max & Median & error  \\
\hline
\usun     &   \kms     &     8.      &     12.         &               10.79   &        0.56  \\ 
 \vsun      &   \kms    &     6.      &     13.         &              11.06   &        0.94  \\ 
 \wsun      &   \kms    &     5.      &      10.        &               7.66   &        0.43  \\ 
 \adisc     &   \kms      &      4.      &     8.    &                  4.59   &        0.14  \\ 
 \bdisc     &   -      &     0.25      &      0.75           &                  0.54   &        0.02  \\ 
 \sigratio$_{\mathrm{\ thin}}$     &   -     &     2.      &     3.2           &                2.58   &        0.06  \\ 
 \suytd     &   \kms       &     28.      &     50.       &            37.91   &        2.17  \\ 
 \swytd     &   \kms       &     25.      &     40.       &            30.96   &        0.96  \\ 
 \hline
 Factors relative to fully self-consistent dynamical model: \\
 \Rsud factor    &   -      &     0.8      &     1.2        &                  1.06   &        0.14  \\ 
 \Rsuytd factor    &   -      &     0.8      &     1.2       &                  0.93   &        0.12  \\ 
 \Rswd factor     &   -    &      0.8      &     1.2          &                 1.01   &        0.11  \\ 
 \Rswytd factor    &   -      &     0.8      &     1.2       &                  0.98   &        0.14  \\ 
 \dvsd factor   & -   &     0.8      &     1.2        &            1.01   &        0.04  \\ 
 \dvsytd factor    &   -     &     0.8      &     1.2        &                  0.98   &        0.08  \\ 
 \Rrhoratiod factor     &   -      &     0.8      &     1.2          &                  1.04   &        0.17  \\ 
 \Rrhoratioytd factor     &   -     &     0.8      &     1.2         &                  0.99   &        0.10  \\ 
 \hline
\end{tabular}
\end{center}
\end{table*}

For each age component, Table~\ref{tab:chapeauDecB7jul}   presents the  values of the derived Shu's DF parameters for the fully self-consistent model. In order to compare these with the findings of similar works, we also show the local values of the radial and vertical velocity dispersions  in the last columns. Also, Table \ref{tab:DM}  presents the fitted parameters of the dark matter halo
 in the final model.

\begin{table*}
\caption{Parameters of the final model for each age component. The first column gives the disc component (index 1 to 7 for the thin disc, `ythd' for the young thick disc). The second column indicates the mean age (Gyr) of the component in the population synthesis model. The third column gives the local density. The fourth to eighth columns give the Shu's parameters corresponding to the radial and vertical velocity dispersions, scale lengths of density, radial dispersion, and vertical dispersion. The last columns indicate the values at the Sun position of \sigR\ and \sigz\ for each age subcomponent for comparison with other studies. The parameters of the first age component of the thin disc are not adjusted in our process because of the lack of axisymmetry.}
\label{tab:chapeauDecB7jul}
\begin{center}
\begin{tabular}{cccccccccc}
\hline
Component &  Mean age &   $\rho_0$ &   $\tilde{\sigma}_R$  &   $\tilde{\sigma}_z$  &  $\tilde{H}_{\rho}$   &   $\tilde{H}_{\sigma_R}$  &   $\tilde{H}_{\sigma_z}$ & $\sigma_{R}(R_0,0)$ &  $\sigma_{z}(R_0,0)$\\
 & Gyr & \Msun\ pc$^{-3}$ & \kms & \kms & pc & pc & pc & \kms  & \kms\\
\hline
1    & 0.075 &  {\it $1.968\times 10^{-3}$}  &   &  & &  &  & {\it 16.7}  & {\it 5.} \\
   2 &   0.575 &    0.598$\times 10^{-2}$ &      9.50 &   3.63 &   2852.0 &    9289.0 &    14508.0 &  14.47 &   8.60\\
   3 &   1.500 &    0.446$\times 10^{-2}$ &    15.49 &   5.92 &    2852.0 &    9979.0 &    13841.0 &  20.54 &  10.86\\
   4 &   2.500 &    0.312$\times 10^{-2}$ &    20.10 &   7.69 &     2852.0 &   8357.0 &    13063.0 &  25.39 &  12.59\\
   5 &   4.000 &    0.556$\times 10^{-2}$ &     25.54 &   9.77 &    2852.0 &   8357.0 &    13063.0 &  31.27 &  14.63\\
   6 &   6.000 &    0.595$\times 10^{-2}$ &    31.41 &  12.01 &     2852.0 &   8357.0 &    13063.0 &  37.92 &  16.80\\
   7 &   8.500 &    0.110$\times 10^{-1}$ &     37.52 &  14.35 &    2852.0 &   8357.0 &    13063.0 &  45.24 &  19.04\\
   ythd &   10.000 &    0.267$\times 10^{-2}$ &     40.18 &  31.86 &     2080.0 &  5986.0 &     6703.0 &  54.66 &  40.53\\
\hline
\end{tabular}
\end{center}
\end{table*}

\begin{table}
\caption{Parameters of the dark matter halo in the final model.  $\rho_c$ is the central mass density, $q_{\mathrm{DM}}$ is the ellipsoidal flattening, $R_{\mathrm{DM}}$ is the core radius.}
\label{tab:DM}
\begin{center}
\begin{tabular}{cccc}
\hline
 & Unit &  Value \\
 \hline
$\rho_c$ & \Msun\ pc$^{-3}$ & 0.202\\
 $q_{\mathrm{DM}}$  &  & 1.054 \\
 $R_{\mathrm{DM}}$ & pc & 3315\\
\hline
\end{tabular}
\end{center}
\end{table}

The determination of some of the parameters suffers from degeneracies as shown in Appendix\ \ref{App:correlations}. One notices for instance the correlation between \adisc\ and \bdisc\, the parameters of the AVR (see Eq.\ref{eq:avr}), although all acceptable combinations of the two produce very similar relations. The thick-disc velocity dispersions are also slightly correlated with the thin-disc velocity dispersions (in the parameters \sigratio, \adisc\ and \bdisc), probably because of the mix of the two populations in many fields. Fields where the thick disc is not polluted by the old thin disc are scarce.  Information about abundances (in particular in $\alpha$ elements) would be necessary in order to improve this. Spectroscopic surveys such as APOGEE, LAMOST, GALAH, or in the near-future \Gaia\  DR3, WEAVE, and 4MOST will be most appropriate to improve the study by better splitting the thick from the thin disc. However, this will be at the expense of more complex selection functions to be accurately estimated and applied to simulations.

\begin{figure*}[h!]

\includegraphics[width=9cm]{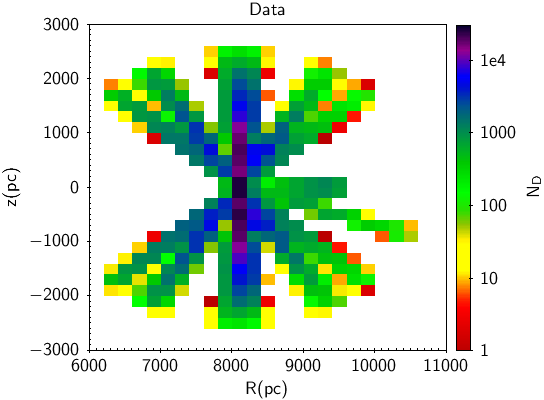}
\includegraphics[width=9cm]{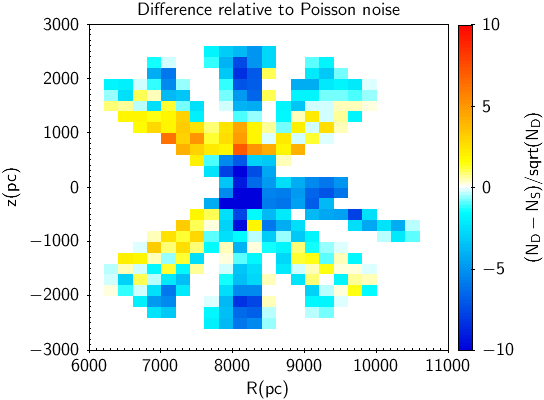}

\includegraphics[width=9cm]{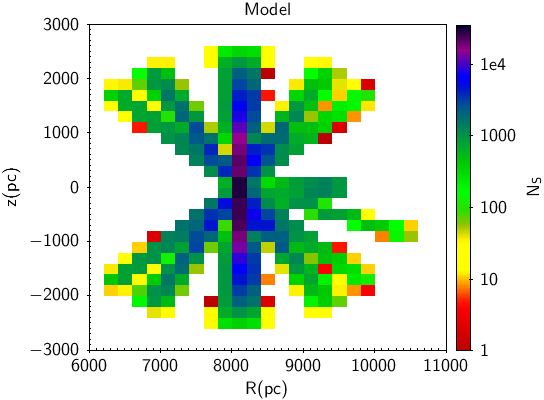}
\includegraphics[width=9cm]{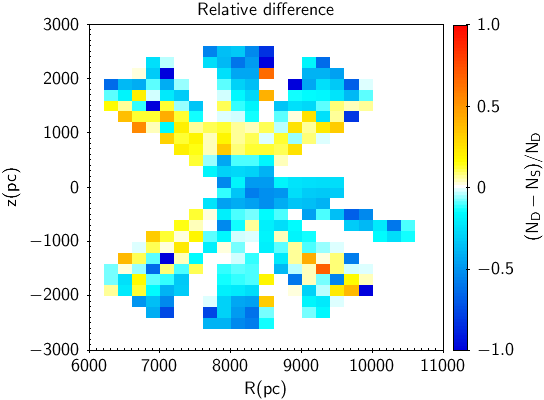}

\caption{
Number density as a function of pseudo $R$ and $z$ (see text) in the selected data set. 
\Gaia\  data N$_D$ (top left) and our final model N$_S$ (bottom left).
Difference between data and model relative to the Poisson noise (N$_D$-N$_S$)/sqrt(N$_D$) 
(top right). 
Relative difference between counts (N$_D$-N$_S$)/N$_D$ (bottom right). The relative difference shows that the model accuracy is generally better than 10\%, apart from the regions showing the vertical wave (see text).
}
\label{fig:Rzgrid-dens}

\end{figure*}

\begin{figure*}[h!]
\begin{center}
\includegraphics[width=9cm]{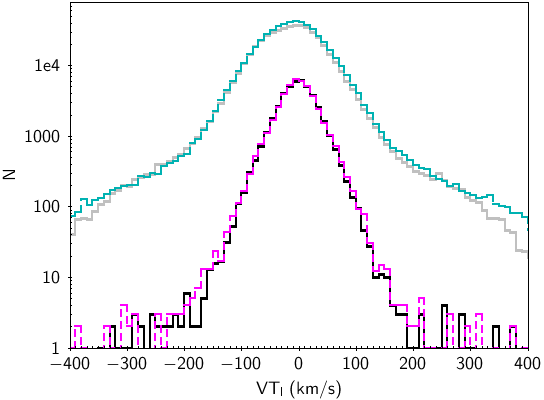}
\includegraphics[width=9cm]{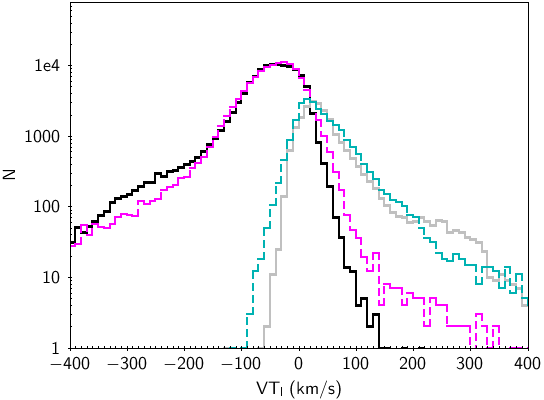}

\includegraphics[width=9cm]{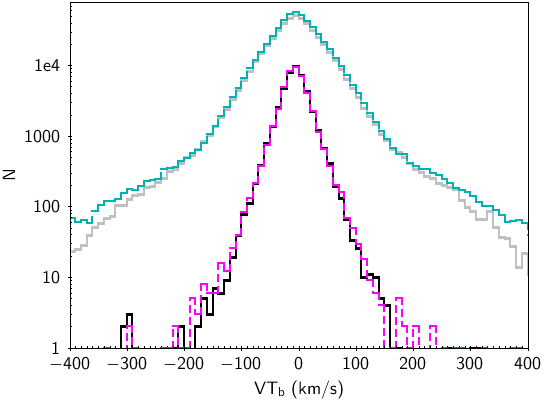}
\includegraphics[width=9cm]{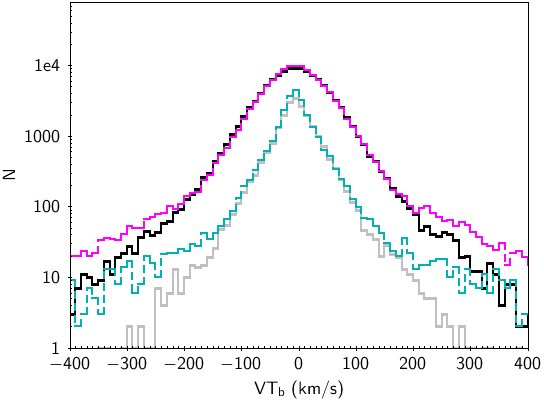}
\caption{
Histograms of the transverse velocities for the local sample and for deep fields, with pseudo-$R$ smaller than 8 kpc or larger than 9 kpc: \VTl\ (top row ), \VTb\ (bottom row). 
{Left column}:  Local data (continuous black line),  local model (magenta dashed line), deep field data (continuous grey line), and  deep field model (dashed cyan line).
{Right column}:  $R<8$ kpc data (continuous black line), $R<8$ kpc model (magenta dashed line), $R>9$ kpc data (continuous grey line), and $R>9$ kpc model (dashed cyan line).} 
\label{fig:hist-VT}
\end{center}
\end{figure*}

In Appendix\ \ref{App:Rz} we show the overall characteristics of the best fit model, notably how the density and kinematics values vary with $R$ and $z$ for each age component. In order to assess the reliability of the fit, we first explored the density distributions obtained and compared them with the data in the Galactocentric coordinates $(R,z)$. We note that the actual star counts are representative of the true density convolved with the selection function which is the same for model and data. Moreover, while in the fit we did not use the distance, but the parallax distribution instead, in these post-fit validation tests, we consider pseudo distances taken as the inverse of the parallax and compute pseudo $(R,z)$ from them. The true $(R,z)$ positions of stars generated in the simulation are disregarded. To allow fair comparisons between model and data, we compute pseudo ($R,z)$ in the simulation from the parallax with errors. In the case of our data set, the error on distance in this way is small because of our selection of relatively bright magnitudes $G<17$, and parallaxes larger than 0.4 mas. 
To alleviate the reading of the paper, we provide figures comparing densities of data and model in the ($R,z)$ space  in Appendix\ \ref{App:Rz}, along with tangential velocity plots of the medians and standard deviations for different sample selections. We present here a summary plot for the density in Fig.\ref{fig:Rzgrid-dens} and velocity histograms in Fig.\ \ref{fig:hist-VT}.

Figure~\ref{fig:Rzgrid-dens} presents the median density on a grid of pseudo $(R,z)$ computed as explained above for the data, the fitted model, and the relative difference, and the difference between the two relative to the Poisson noise $\sqrt{N_D}$. 
The relative difference has a mean of  $-0.14$, a median of $-0.09,$ and a standard deviation of $0.37$. We notice some systematic errors, particularly along the $z$ axis where the vertical wave at the Sun noted by \cite{2019MNRAS.482.1417B} and \cite{2020A&A...643A..75S} appears clearly, particularly towards the north with an excess in the data at $z\approx$ 900 pc, and deficits around 400 pc and 2 kpc.

The excess in the model near the plane is mainly due to an excess in bright and young stars. We noticed that the model is in excess of massive stars (by about 50\%) but they represent only 4\% of the selected sample and should not impact the global fit. This could be due to either the IMF at high masses (mass larger than 1.5 \Msun), which would have an overly shallow slope (the assumed IMF slope at high mass is $\alpha=2.5$), or the recent star formation rate (ages below 2 Gyr). We performed several tests to verify this assertion by changing the IMF slope. The resulting dynamical fit was exactly the same. The massive stars do not constitute a major component of the stellar mass. Therefore, this is not a problem for the dynamics. However, we shall reconsider the IMF and SFH of the model in the near future using the most recent Gaia DR3 data.

We finally present comparisons of histograms of \VTl\ and \VTb\ for the local and deep fields, and for different pseudo $R$ ranges in Figure~\ref{fig:hist-VT}. The overall agreement clearly appears here. But the sample is also sufficiently large to show some sticking points, especially in the wings of the distributions. We point out that, at \VTl > 200 \kms\ on the outer side (\Rgal > 9 kpc, cyan line in top-right panel) and at \VTl < $-$ 150 \kms\ on the inner side (\Rgal < 8 kpc, magenta line in top-right panel), the model presents a lack of stars in the wings of the distribution. In particular, our model does not include the Gaia-Enceladus component which contributes to the wings at $V_{tot}$>200 \kms\ as shown in \cite{2018A&A...616A..10G}. However, globally the model presents a slight excess of the halo (or may be of the old thick disc) population in the wings, a problem that will be handled in the near future.  In our sample, this concerns only a small portion of the stars in the wings, and so it should not bias our results concerning the thin and thick discs.

\section{Characteristics of the new dynamical model}
\label{Sect:newmodel}
\subsection{Comparison with previous BGM}

\begin{figure}[h!]
\begin{center}
\includegraphics[width=9cm]{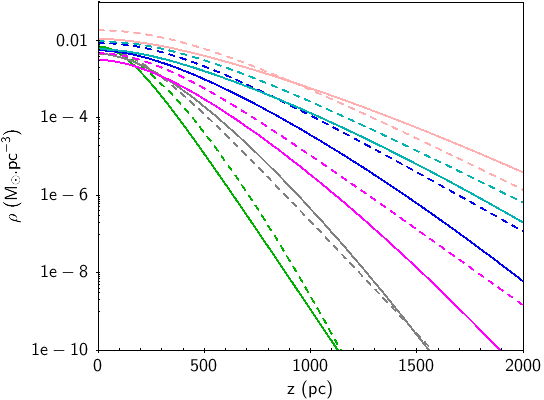}
\caption{Comparison of density-law variation with $z$ between the previous BGM model (dashed lines) and the new self-consistent model (solid lines) at \Rsun\ for the thin-disc components. 
Age component number 2 (green, age 0.15-1 Gyr), 3 (grey, age 1-2 Gyr), 4 (magenta, age 2-3 Gyr), 5 (blue, age 3-5 Gyr), 6 (cyan, age 5-7 Gyr), and 7 (pale pink, age 7-10 Gyr). }
\label{fig:dens-comp}
\end{center}
\end{figure}

The density laws constrained by the dynamics are significantly different from those of the previous model \citep{Robin2003} which followed Einasto laws and applied the self-consistency principle at \Rsun\ as in \cite{Bienayme1987a} assuming the age--velocity relation of \cite{Gomez1997}. Figure~\ref{fig:dens-comp} shows the variation in density as a function of \zgal\,  at the Solar Galactocentric radius for the different age components for the Mev2011 version and the new dynamical model. The difference is small for ages younger than 2 Gyr and ages older than 5 Gyr, but is significant for intermediate ages.
 
\subsection{Outer Galaxy flare}

\begin{figure}[h!]
 \begin{center}
\includegraphics[width=9cm]{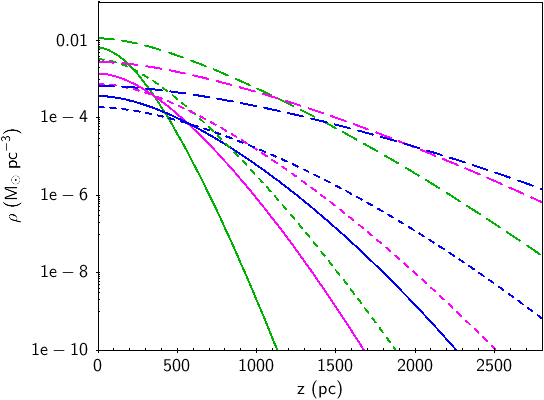}
\includegraphics[width=9cm]{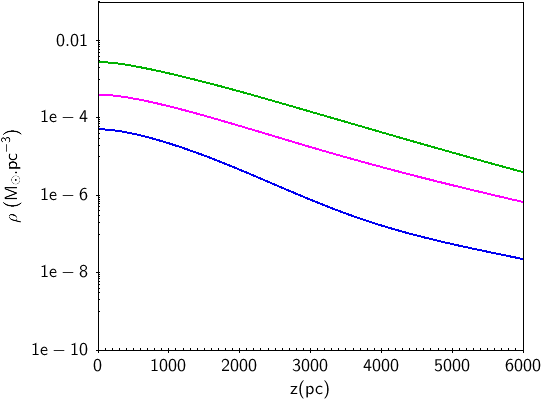}
\caption{Density laws as a function of $z$ for three different Galactocentric radii ,$R$=8, 12, and 16 kpc in green, magenta, and blue, respectively. Top panel: For thin disc components 2 (solid line), 4 (short dashed line), and 7 (long dashed line). Bottom panel: For young thick disc.}.
\label{fig:flare}
\end{center}
\end{figure}

As a distinctive feature, the model naturally produces a flare in the thin disc, not present in the case of the thick disc. This is seen in Fig.\ref{fig:Rz-dens} on the intervals of the iso-density contours which appear to be wider at high $R$ than in the Solar neighbourhood. Figure~\ref{fig:flare} illustrates the flare in the thin disc (top panel) by showing the vertical decrease for age components 2, 4, and 7 of the thin disc, at a Galactocentric radius of 8, 12, and 16 kpc. The apparent scale height increases clearly with $R$ in all thin-disc age components. Considering the thick disc, the bottom panel of Fig.~\ref{fig:flare}  shows the opposite behaviour,  where a smaller scale height is seen when $R$ increases from 8 to 16 kpc.
The thick disc is not flaring, but rather on the contrary it shrinks when going to the outer Galaxy. This clearly indicates a different dynamical origin for this population compared to that of the thin disc. 

\subsection{Age--velocity dispersion relation}

\begin{figure}[h!]
    \centering
    \includegraphics[width=9cm]{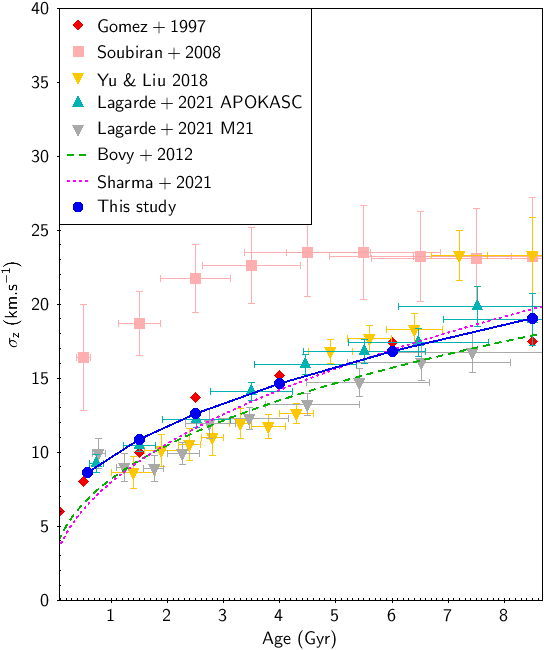}
    \caption{Age--vertical velocity dispersion relation at the Solar position. Red diamonds: \cite{Gomez1997} from Hipparcos. Yellow triangles: \cite{2018MNRAS.475.1093Y} from LAMOST red giants with [Fe/H]$>$-0.2. Pink squares: \cite{2008A&A...480...91S} from red clump giants. Cyan triangles : Thin disc population from \cite{2021A&A...654A..13L} with ages from \cite{2021A&A...645A..85M}(M21). Dark grey triangles: Thin disc population from \cite{2021A&A...654A..13L} with ages from APOKASC \citep{2018ApJS..239...32P}. Green dashed line: Relation from the model of \cite{Bovy2012a}. Magenta dotted line: Relation from  \cite{2021MNRAS.506.1761S}. Blue filled circles: This study. 
    }
    \label{fig:avr}
\end{figure}

The AVR is an important result of our fit
presented in Fig.~\ref{fig:avr}. It shows a significant increase with age, although we do not see a net plateau, as was found by \cite{Gomez1997} for example. We further discuss this result in the context of the related literature in Sect.\ \ref{Sect:discussion}.  Generally, the AVR is given at the solar position, while in our case the velocity dispersions vary both in $R$ and $z$ as can be seen in Fig.\ \ref{fig:Rz-dens}, and vary differently from one thin disc component to another.

\subsection{Dichotomy between thin and thick discs}
\label{Sect:dichotomy}
It is interesting to compare the characteristics of the thick disc generated self-consistently with those of the old thin disc, as this can shed light on their formation histories. As pointed out above, the thick disc does not present any flare while there is strong flaring in the thin disc population. This is in line with the results of many spectroscopic surveys, which show that the thick disc, when selected by high-alpha abundances, has a density that drops fast in the outer Galaxy while the low-alpha thin disc remains prominent and is even found at higher $z$ in this region \citep[][among others]{Hayden2015ApJ...808..132H}.

\begin{figure*}[h!]
 \begin{center}
\includegraphics[width=9cm]{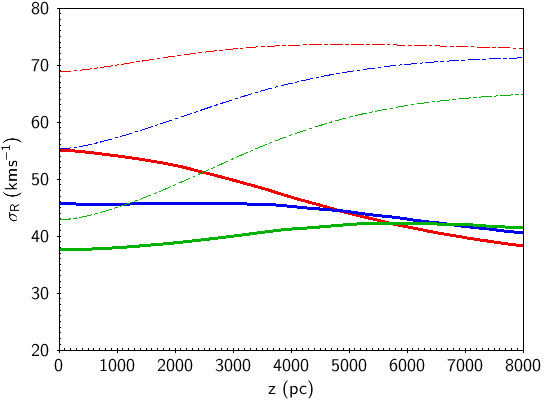}
\includegraphics[width=9cm]{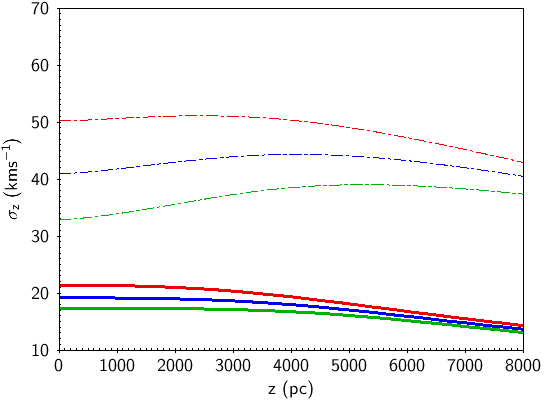}
\includegraphics[width=9cm]{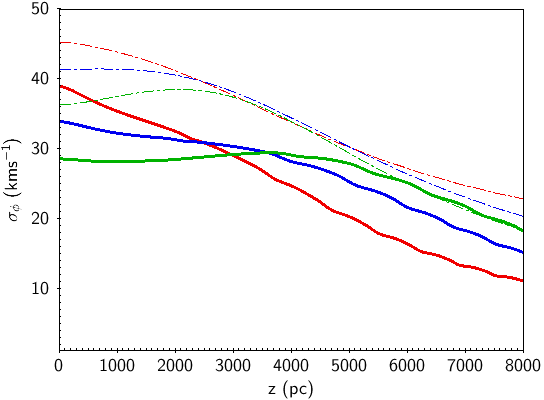}
\includegraphics[width=9cm]{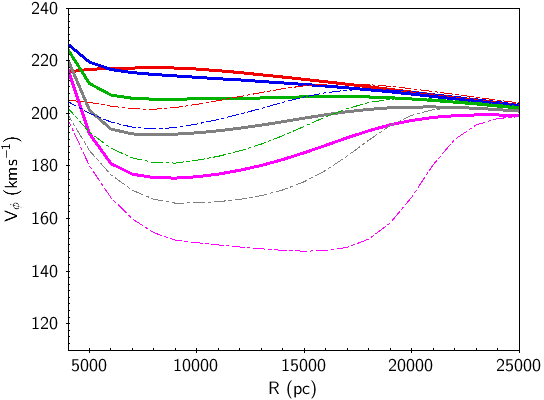}
\caption{\sigR, $\sigma_\phi$, and \sigz\ as a function of $z$ for three different Galactocentric radii $R$=6, 8, and 10 kpc, in red, blue, and green respectively, for the thin-disc component 7 (solid line) and the young thick disc (dotted dashed). Bottom right: Mean azimuthal velocity as a function of $R$ for different $z$ (red: $z$=0; blue: $z$=1 kpc; green: $z$=2 kpc; grey: $z$=3 kpc; magenta: $z$=4 kpc) for the thin disc component 7 (solid lines) and the young thick disc (dotted dashed lines).}
\label{fig:sig-age7-8}
\end{center}
\end{figure*}

Locally, the \sigz\ of the young thick disc is well above that of the thin disc, with a value of 40.5 \kms. Its value is about twice that of the old thin disc at the age of 10 Gyr.
If one carefully looks at the kinematic distributions in the $(R,z)$ plane (Appendix\ \ref{App:Rz}), some other striking differences appear between the oldest thin disc (age component 7 with age range 7-10 Gyr, in column 6 in Fig.\ \ref{fig:Rz-dens}) and the young thick disc in column 7. This can be seen even more clearly in Fig.\ \ref{fig:sig-age7-8}.

As shown in Fig.\ \ref{fig:sig-age7-8}, the  difference in kinematics between old thin disc and thick disc populations is mainly in the vertical velocity dispersions, while the difference is weaker when we consider the radial velocity dispersion or the tilt angle of the ellipsoid. The tilt seems to change smoothly from one component to another (see Fig.\ \ref{fig:Rz-dens} last row). On the other hand, the mean rotation velocities are also significantly different, that is, the asymmetric drift in the thick disc is about twice that in the old thin disc at a given $z$ and at the solar Galactocentric radius. However, in the thick disc $V_\phi$ decreases most significantly when going at large $R$ and $z$ (as shown in magenta curves in the bottom right corner of Fig.\ \ref{fig:sig-age7-8}).

{
Our model shows that the dichotomy between the old thin disc and the thick disc clearly appears in the DFs of our self-consistent model. If one assumes that our thick disc corresponds to the $\alpha$-rich population, this nicely explains previous results showing that the thick disc defined as such has a density that drops in the outer Galaxy.

\subsection{Mass distribution, and radial and vertical forces}

Our model provides new measurements of the mass distribution of our Galaxy and its gravitational forces. The observational constraints are based on measurements of the Galactic rotation curve and the  local dynamical measurement of the dark matter mass density (from the Oort limit and $K_z$ measurements). On the other hand, the constraints on the stellar counts also provide a measurement of the stellar mass. Finally, taking into account the stellar kinematics brings an additional constraint by considerably reducing the number of free parameters while ensuring the dynamical consistency of the model by fitting  stationary distributions.

In this study, we adopted a dark halo with a local density of 0.010 \Msun pc$^{-3}$ \citep{2020A&A...643A..75S}  smaller than that adopted by \cite{Bienayme2015A&A...581A.123B}. This value is similar to those used by the previously cited works. A small change of 10 percent of this adopted value would only change the total local density and vertical forces  by 1 percent. Therefore it would induce a very small change in the computed vertical stellar density distributions. Moreover, increasing the local dark matter density by ten percent is nearly equivalent here to the  flattening the dark halo by about ten percent, thus recovering almost the same  Galactic rotation curve. This implies that the uncertainty on the local dark matter has only a weak impact on our modelling. We can conclude that, assuming a local density of 0.010 \Msun pc$^{-3}$, the dark matter component is nearly spherical, $q\sim 1$ at least in the inner 20 kpc of the Galaxy.

The vertical force as a function of $R$ is presented in Figure~\ref{fig:kz} for $z$=0.5, 1.1, and 2 kpc, and as a function of $z$ for three different Galactic radii $R$=6, 8.1, and 10.2 kpc. \cite{2022MNRAS.510.2242W} also performed a global dynamical model of the Galaxy and used various constraints, relying on a simplified stellar disc mass model. They obtained results close to ours (see Figure~\ref{fig:kz}) for the vertical force distribution at $z=1.1$ kpc \citep[see figure 7 in][]{2022MNRAS.510.2242W} and for $R>8$ kpc,
which is also quite comparable to the measurements of \cite{2013ApJ...779..115B}. We emphasise that their parameterisation of the stellar discs is different from ours, with a shorter density scale length, and this explains the difference in $K_z$ forces for Galactic radii $R<$ 8 kpc. 

\cite{2021ApJ...916..112N} developed a  dynamical model of the Milky Way based on 6D phase space data from APOGEE and \Gaia\  eDR3. The analysis of these authors is based on solving the Jeans equations  to model the stellar velocities and dispersions. They obtain Galactic mass distributions in the range of Galactic radii from 5 to 19 kpc  with values of $K_{z_{1.1}}(R)$ \cite[see figure 7 in][]{2021ApJ...916..112N}, which is also similar to those of \cite{2022MNRAS.510.2242W}.

\begin{figure}[h!]
\begin{center}
\includegraphics[width=9cm]{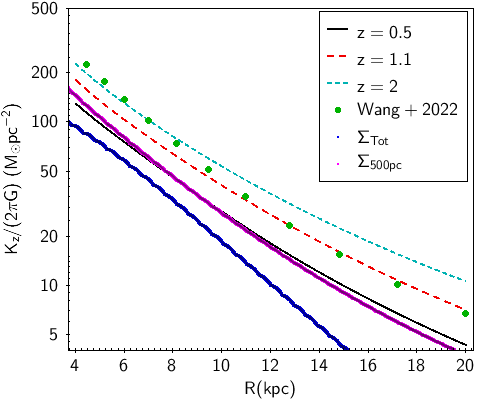}
\includegraphics[width=9cm]{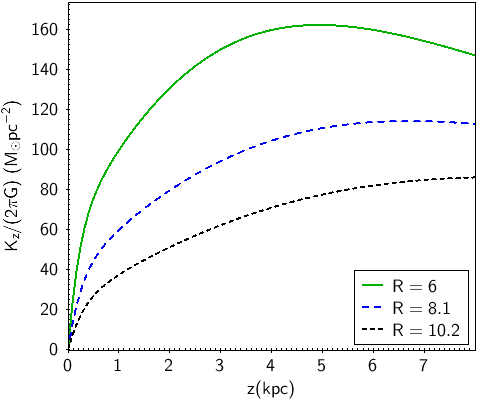}
\caption{Galactic vertical force. Top panel: $K_z$ forces versus Galactic radii at $z$=0.5 (black thin solid line), 1.1 (red dashed), and 2 kpc (cyan dotted) from the Galactic plane (red). Green dots are the $K_z$ force at 1.1 kpc from \cite{2022MNRAS.510.2242W}. The blue thick line is the total surface mass density of stellar discs. The magenta line is the total (stellar+ISM+dark matter) density below $|z|=500$ pc that can be compared to the $K_z$ force at 500 pc. Bottom panel: $K_z$ forces versus $z$ at three Galactic radii $R$=6 (green solid line), 8.1 (blue dashed), and 10.2 kpc (black dotted line). } 
\label{fig:kz}
\end{center}
\end{figure}

As for the mass distribution of the stellar discs, we find $\Sigma_*(R_0)$=32.2\ \Msun pc$^{-2}$ for the surface
mass density of all discs at the solar position. This makes $\Sigma_{baryon}(R_0)$=43.2\,\Msun pc$^{-2}$  lower than the value $\Sigma_{baryon}(R_0)$=54.2\,\Msun pc$^{-2}$ \citep{2014JPhG...41f3101R} and close to the value of $\Sigma_{baryon}(R_0)=48.7$\,\Msun pc$^{-2}$  \cite[see table 2 in][]{2006MNRAS.372.1149F}. Figure \ref{fig:kz} shows the distribution $\Sigma_*(R)$ as a function of the Galactocentric radius. The distribution is quasi-exponential with a scale  length  of $R_\Sigma =$ 3.48 kpc  in the interval $R=6$ to 10 kpc for the sum of all stellar disc components. This value is significantly larger than the value of 2.5 kpc derived from the $K_z$ measurements by \cite{2013ApJ...779..115B}.

 We must note that, for the thin discs, the effective scale lengths vary and increase significantly with $z$, thus the resulting surface density scale length for thin disc components is about $R_\Sigma =$ 3.9 kpc. For the young thick disc, it is $R_\Sigma =$ 2.31 kpc.  

\section{Discussion}
\label{Sect:discussion}
Our method allows us to derive a self-consistent model for the  stellar populations with densities consistent with the velocities and the Galactic gravitational potential, and is able to reproduce the corresponding \Gaia\  data in an extended volume around the solar position.
This is a very useful step towards ensuring that the Milky Way potential we have is realistic. 

Our model is not strictly comparable with analyses where the density and velocity distributions are derived from observational samples. Those analyses are generally biased by the sample selection, while our model itself is not. These studies present different ways to circumvent the problems of biases, including partial modelling and different assumptions. Furthermore, our model and methodology  provide a safer way to compare data and models in the observational space and to reproduce the selection function as
accurately as possible, as we do in the present work.

Therefore, a robust comparison between different data samples should be done by simulating observed samples one by one with a population synthesis model, accounting for observational errors, as we do for Gaia data in the present work. In this section, we estimate the similarity between the conclusions provided by our model regarding the kinematics of the stellar populations of the Milky Way and those of other models and studies. We particularly discuss the question of the solar motion, disc kinematics, and density laws.

\subsection{Solar motion}

Our analysis leads to a solar motion of \usun=10.79 $\pm$ 0.56 \kms, \vsun=11.06 $\pm$ 0.94 \kms, and \wsun=7.66 $\pm$ 0.43 \kms with respect to the LSR.
\cite{2021MNRAS.504..199W} presented a summary of most references in the literature concerning the solar motion, showing that  \wsun\ has a consensual value of about 7 \kms\ in recent works, while the accepted range
of  \usun\ still covers 7 to 12 \kms, and \vsun\ shows even more conflicting values, ranging between 1 and more than 20 \kms. From LAMOST data and \Gaia\  DR2, using a local sample of A-type stars, \cite{2021MNRAS.504..199W}
find the mean solar motion to be $(11.69\pm0.68,10.16\pm0.51,7.67\pm0.10) $ \kms. We note that even if the solar motion with respect to the LSR should not depend on the selected sample, their determination does depend on the sample selection. Moreover, they assume no variation of the velocity ellipsoid with Galactic position. For example, using stars at distances up to 1 kpc, \cite{2021MNRAS.504..199W} find a \usun\ of 9.79 while \vsun\ decreases to 8.50, at about 3 sigma of their finally claimed values. However, their result on average and according to the error bars remains in good agreement with ours for the three components of solar motion.
\\
\cite{Dehnen1998MNRAS.298..387D}, 
 from Hipparcos data, found values of $(10.00\pm 0.36, 5.25\pm 0.62, 7.17\pm 0.38)$ showing very good agreement at less than 1 sigma with our results for \usun\ and \wsun. Their \vsun\ value was already shown to deviate from that of \cite{10.1111/j.1365-2966.2010.16253.x} who found ($11.1^{+.69}_{-0.75}$, $12.24^{+0.47}_{-0.47}$, $7.25^{+0.37}_{-0.36}$) \kms. The latter present a detailed chemo-dynamical model to compute solar motion. They show that using a large range of stellar types, they are able to fit a model where the \vsun\ does not depend on metallicity.
 Our model differs not only by the dynamical modelling, but also by the data used (Hipparcos data in their case) even though the solar motion found in both cases is very similar. \cite{10.1111/j.1365-2966.2010.16253.x} already suggested that the results may vary with the sample due to the substructures in the local data and that a final answer for \vsun\ will vary when more distant and accurate data become available. Non-axisymmetry and streaming motions may also play a part in the problem; \cite{2015ApJ...800...83B} evaluated these latter at the level of 11 \kms\ on scales of 2.5 kpc.
 
 \cite{2018A&A...614A..63S} 
 determined constraints on the solar motion together with the local rotation curve from RAVE and SEGUE data. These authors used the model of \cite{Golubov2013A&A...557A..92G} to correct for the asymmetric drift and found \vsun\ to be 4.47$\pm$ 0.8 \kms\ which is smaller than our value and that of \cite{10.1111/j.1365-2966.2010.16253.x} 
 and is closer to that of \cite{Dehnen1998MNRAS.298..387D}.  
 This value, as ours, depends on the Galactic model used. Surprisingly, the rotation curve found
 by \cite{2018A&A...614A..63S} is flat or rising at the Sun position, and also depends on metallicity. We believe that this is an indication that there is bias due to the selected sample, or to imperfect correction of the asymmetric drift that produces this dependency on metallicity (see also \cite{10.1111/j.1365-2966.2010.16253.x}).
 
 Using \Gaia\  DR1 astrometry and RAVE spectroscopy in \cite{2017A&A...605A...1R}, we found a slightly different value for \usun\ of 13.2 \kms compared to 10.65 \kms\ here, and a significantly smaller value of 1 \kms\ for \vsun\ compared to 11 \kms\ here}.  In the former study, the potential is slightly different and the stellar densities are not self-consistently computed; the kinematical scale lengths and variations of the asymmetric drift with $R$ and $z$ are also different and fixed a priori. \cite{2017A&A...605A...1R}
 used RAVE data together with \Gaia\ DR1, which covered a limited range in distance and are less accurate than \Gaia\  eDR3. The disagreement with the present study is mainly due to the overall shape of the potential used at that time which has a significant impact on the value of the asymmetric drift, as well as on the mean circular velocity at the Sun position. In the present study, we performed full determination of self-consistent density and kinematics and compared them with a much larger data sample, giving more confidence in the result.
 
 It is worth mentioning that we obtained consistent values for the solar motion, at the level of 1.5 $\sigma$, using either the local sample alone, or the combination of it with our deep fields. The result is robust with the sample selection, most probably because our model reliably accounts for the variations of the circular velocity of the stars with their age and position in the Galaxy.

\subsection{Thin- and thick-disc kinematics}

\paragraph{Thin disc AVR.}
Our study allowed us to determine the thin-disc kinematics, fitting the age--vertical velocity dispersion relation, the vertical-to-radial-velocity dispersion ratio, and the kinematic and density scale lengths. For the former, we find very consistent values with the literature, with vertical velocity dispersion varying from 10 to 20 \kms\ for stars of 0.1 to 10 Gyr near the Sun (see Fig.~\ref{fig:avr}). Most importantly, we compare here an AVR at the Sun with AVR on samples that can cover a wide range of distances from the Sun (especially for giants). Ages are also difficult to determine from observations as absolute values. Our AVR shows values that are consistent with the \cite{Gomez1997} AVR from the Hipparcos sample, although the AVR of these latter authors exhibited a saturation of the vertical velocity dispersion for the old thin disc at 15-17 \kms, while we find a slightly higher maximum value for the thin disc at the level of 19 \kms. \cite{2021A&A...654A..13L}, from \Kepler\ giants, found vertical velocity dispersions slightly below ours for young stars when ages are estimated from \cite{2021A&A...645A..85M} (M21 in the figure), although with ages from APOKASC the agreement is relatively good. Comparing with \cite{2018MNRAS.475.1093Y} who studied the LAMOST sample of red giants with thin disc metallicities ([Fe/H] $>$ -0.2 dex), the agreement is also good for ages of above 5 Gyr, but for young stars these latter authors obtain a slightly lower velocity dispersion. \cite{2008A&A...480...91S} on the contrary found higher values at all ages from another red clump sample. For comparison,  in Fig.~\ref{fig:avr} we also plot age--velocity dispersion relations obtained by \cite{Bovy2012a} and \cite{2021MNRAS.506.1761S}. 

\cite{Dehnen1998MNRAS.298..387D} analysis of the local kinematics gives consistent values for the $\sigma_R/\sigma_z$ of 2.2 and a radial velocity dispersion of 38 \kms\ for the oldest thin disc stars, which is slightly lower than our value of 45 \kms. Their ratio of scale length $H_\sigma$/$H_\rho$ is 3 to 3.5 while we find 2.9. We find a $\sigma_R/\sigma_z$ of 2.38 at the Sun position, but the value varies with $R$ and $z$ to reach 1.6 for example at $R=15$ kpc and $z$=0. 
\sigR/\sigz\ values of close to 2 are rather frequent in the literature, and measurements are taken quite often in the solar vicinity, such as in 
\cite{2007A&A...475..519H} from the Geneva Copenhagen survey, \cite{Aumer2009MNRAS.397.1286A} from Hipparcos data, \cite{2018MNRAS.475.1093Y} from LAMOST, or \cite{2017A&A...602A..67A} from RAVE and \Gaia\  DR1, among others. 

Notably, \cite{2018MNRAS.481.4093S} 
used \Gaia\  DR2 complemented by spectroscopic surveys to determine ages for 3 million stars and found variation of the age--velocity dispersion relation with Galactocentric radius. These authors used a Bayesian approach to determine distances from parallaxes assuming priors, and isochrones to determine ages. They found that the radial velocity dispersion $\sigma_R$ as a function of Galactocentric radius continues to decline at $R>$ 10 kpc, while the $\sigma_z$ stops declining around the Sun, and then slightly increases at larger $R$. Our model does not produce this up-turn in \sigz\ at large radii in the Galactic plane. However, our gradient flattens at larger $z$ such that the values of \sigz\ are quite different in the plane from those at distances from the plane. 
 \cite{2018MNRAS.481.4093S} stressed that,  age uncertainties increasing with age can bias the overall estimation of the age--velocity dispersion relation. In our method, where no ages are assumed for the stars, we avoid this kind of bias.
 
\cite{2018A&A...616A..11G} 
studied \Gaia\  DR2 and presented the same flattening of the vertical \sigz\ at larger $R$ (see their figure 18), which is compatible with our result. Contrarily to \cite{2018MNRAS.481.4093S}, \cite{2018A&A...616A..11G} did not see an upturn in \sigR\ or \sigz\ at large $R$, at least up to 14 kpc, and their sample is less biased by the selection function.

\cite{2021MNRAS.506.1761S} 
present a complex analysis of the kinematics and its dependency on position, age, and metallicity from multiple surveys (GALAH, LAMOST, APOGEE, the NASA \Kepler\ and K2 missions, and \Gaia\  DR2).
They find an AVR following a power law with an exponent of $\beta\ =0.441 \pm $0.007  for \sigz that is slightly steeper than ours. However, their fit applies globally to the thin and thick discs, while our power law index of 0.54$\pm$ 0.02 applies only to the thin disc, and our thick disc has a significantly higher dispersion of 40.53 \kms, whatever its age. 
\cite{2021A&A...647A..39S}, in their overall fit of the Just and Jahreiss model to \Gaia\  DR2, found a slope $\beta\ = 0.41 \pm 0.04 $ for the  AVR for the thin disc, and a $\sigma_z$ of 43 \kms\ for the thick disc, in excellent agreement with ours.

In many observational studies, the velocity dispersion increases with height from the plane. In our model, this is not the case for the thin disc when considering each isothermal (mono-age) population separately. However, this gradient is naturally created when a realistic mix of different populations is considered. In observational samples, such as in \cite{2019MNRAS.489..176M} or \cite{2021MNRAS.506.1761S}, this is produced by the mix of populations, where older stars (with higher dispersions) dominate at higher \zgal.

\cite{2019MNRAS.489..176M} 
studied a large sample of 65000 stars from APOGEE DR14 and \Gaia\  DR2, covering a wide range of distances $z<2$ kpc and 4$<R<$13 kpc. These authors determined ages and used abundances to distinguish low-$\alpha$ and high-$\alpha$ sequences corresponding to thin and thick discs, respectively. Among other interesting results, they see very little variation of \sigR\ and \sigz\ with $z$ in each mono-age population. This is also what is seen in our model (Appendix\ \ref{App:Rz} Fig.~\ref{fig:Rz-dens}).
Our maps of \sigR\ and \sigz\ show very little change with $z$ over a wide range of $R$.
In our model, to see a clear change of dispersion with $z$ it is necessary to go to large distances from the plane, in regions where the stellar densities are very small.
\cite{2019MNRAS.489..176M} also find very long kinematic scale lengths for the thin disc but with a large uncertainty $R_{\sigma_{R}}$ = 15$^{+11}_{-4}$ kpc and $H_{\sigma_{z}}$ = 16$^{+19}_{-5}$ kpc, still compatible with ours.

Concerning the AVR, \cite{2019MNRAS.489..176M} modelled it with a power law index $\beta_z $ of about 0.5 for the thin disc and 0.45 for metal-poor stars, compatible with the value of 0.54 that we obtained in this study for the thin disc. On the other hand, they claim a velocity dispersion ratio 
\sigz/\sigR=0.64$\pm$0.04 for old stars. We find values varying from 0.4 to 0.7 for the thin disc, and around 0.74 for the thick disc (not distinguishing their age) with very little variation with $R$ and $z$  in this case, in very good agreement with their observations.

\cite{2021A&A...654A..13L} 
analysed the \Kepler\ sample to determine the characteristics in age, metallicity, and $\alpha$ abundances of the thin and thick discs. The selected sample is roughly at solar Galactocentric radius. We might therefore expect the kinematics to be close to those of the solar neighbourhood (assuming axisymmetry). For the thin disc, these latter authors derive a vertical velocity dispersion that depends on age, varying from 10 \kms\ at 1 Gyr to 18 at 8 Gyr and a velocity ratio $\sigma_R/\sigma_z$  of about 2.5. They found lower values (35 \kms) than ours for $\sigma_R$ and consistent values for $\sigma_z$.  
\cite{2021A&A...654A..13L} also find that the dependency of the velocity dispersion depends on metallicity, which we are not considering here. This merits further investigation. 

Concerning the radial scale lengths, \cite{2021MNRAS.506.1761S} underlined that the dispersion falls off exponentially as a function of guiding radius, 
which is also a result that is completely in line with our model. However, these authors found that the scale length of \sigR\ ($H_{\sigma_R}$) is larger than that of $\sigma_z$ while we find the contrary. 

In conclusion, our vertical AVR is compatible with most previous determinations for old stars, and is slightly higher for young stars, but within the error bars. This slight discrepancy could be investigated further to explore whether or not it could be due to the fact that, here, we compare the velocity dispersions at the Sun position, while various studies cover a wide range of positions depending on the selection functions. 

\paragraph{Thick disc kinematics. }

In our model, the vertical velocity dispersion in the young thick disc slightly increases with \zgal\ at $R_0$ and decreases when \Rgal\ increases (see Appendix\ \ref{App:Rz},  fig.\ref{fig:Rz-dens}). 
In a study of the \Kepler\ field, \cite{2021A&A...654A..13L} 
found that the thick disc vertical velocity dispersion depends on alpha abundance and metallicity and ranges between 25 and 40 \kms\ along the alpha sequence from low to high $[\alpha/\mathrm{Fe}]$, while we found 40.53 \kms\ at the solar position. The \sigR\ found by \cite{2021A&A...654A..13L} ranges between 42 and 55 \kms, in good agreement with our value of 54.66 \kms\ locally.
 In order to study the interface between thin and thick discs, the thick disc population of these latter authors was divided into a metal-rich (i.e. with metallicity close to solar) and a metal-poor component, the latter being close to our young thick disc component, while the metal-rich one is not considered here (due to lack of metallicity information). In order to better compare our result in the \Kepler\ field, stellar abundances have to be considered, which is beyond the scope of this paper. Interestingly, the metal-poor thick disc velocity dispersion found by \cite{2021A&A...654A..13L} varies with metallicity and $[\alpha/\mathrm{Fe}]$ but the variation with age is not monotonic and significantly differs from the thin disc behaviour. This clearly indicates a different formation history for the thick disc and the thin disc. This difference in scenario of formation between the thick and thin discs was pointed out in several chemo-kinematical studies, such as \cite{2019MNRAS.489..176M} 
who found a nearly constant ${\sigma}_R/{\sigma_z}$ at a level of 1.56 for the high-$\alpha$ population (thick disc), very close to our model which also shows a very smooth distribution of this ratio nearly independent of \Rgal\ and \zgal\ (Appendix\ \ref{App:Rz} Fig.\ref{fig:Rz-dens} in penultimate row). This dichotomy between thin and thick discs is in line with our model, as already discussed in Sect.\ref{Sect:dichotomy}.

One of the difficulties of comparing our results with others is due to the fact that we have not considered the abundances in this study. This is appropriate when separating the thin and thick discs by $\alpha$-element abundances rather than by age (because they overlap) and/or position (which would refer to the geometrical thick disc instead). Therefore, combining this new dynamical model to the population synthesis approach to simulate abundance surveys will be a very efficient tool to test chemo-dynamical scenarios for the formation of the thin and thick discs. Despite lacking abundances, we pointed out an abrupt transition between thin and thick discs.
Even though a more detailed study by simulating the spectroscopic surveys would be needed to compare our model with chemically selected samples, these preliminary comparisons give us confidence in the reliability of our model to reproduce the velocity dispersions of observed mono-age populations.

\subsection{Density laws}

The new density laws slightly differ from the Einasto laws used in previous BGM, especially as a function of \zgal, as seen in Figure~\ref{fig:dens-comp}, but also radially. The density scale lengths of the thin disc are larger in the plane, but vary greatly with $z$ and the new model exhibits a significant flare in the thin disc. 
 There is a slight difference in $\rho(z)$ for intermediate discs (between 2 and 5 Gyr). 
 As the thick disc scale length $H_\rho$ is shorter than that of the thin disc, the thick disc density, as seen in Appendix Figure~\ref{fig:Rz-dens}, is dense only in the inner Galaxy, while the density drops significantly in the outskirts. This is in line with studies who find larger  exponential scale lengths in the thin disc than in the thick disc \citep[among others]{2013ApJ...779..115B,2014A&A...564A.115A,2014A&A...569A..13R,2018MNRAS.479..211M,2018A&A...614A..63S}. On the contrary, \cite{Golubov2013A&A...557A..92G} found larger scale lengths for the metal-poor population, such as the thick disc,
 from RAVE data (3 kpc, while 1.8 kpc for the thin disc). This last surprising result would be difficult to reconcile with the paucity of the high-$\alpha$ thick-disc stars seen in the outer Galaxy \citep{Hayden2015ApJ...808..132H}.

Our model produces a significant flare in the outskirts of the thin disc (see Appendix\ \ref{App:Rz}, fig.\ref{fig:Rz-dens} and Fig.~\ref{fig:flare}). In practice, the isodensity contour intervals are larger at high $R$ than in the solar neighbourhood. This is notable for young ages (components 2 to 5). In older thin-disc components, the flare is flattening at $R>15$ kpc but this is in a region where the density is very low. 
Such flaring has already been identified in many studies, for instance in \cite{2017A&A...602A..67A}.
Notably, \cite{2020A&A...637A..96C} 
analysed \Gaia\  DR2 to determine the structures in the anticentre and find evidence for asymmetric flaring (different above and below the plane). It is beyond the scope of this paper to explore the features in the outer Galaxy, where a wider selection of data in this region is required to separate the axisymmetric components from the non-axisymmetric ones and from the substructures, which can result from old mergers and gravitational response to perturbations by satellites.

\subsection{Comparison with other models}

The model we present here combines several elements that are rarely found together in other Galactic models. The dynamical coherence links the gravitational potential to all the Galactic components. The stellar distribution functions, density and kinematics, are themselves consistent with the potential. Moreover, the stellar populations are described using the most recent evolutionary tracks. 

\cite{2012MNRAS.426.1324B}  built a $f(J)$ axisymmetric Galactic model based on actions, an approach that also uses St\"ackel potentials as we have developed in this study. The stellar distribution functions of this latter author are stationary as in our model.  He derived a realistic potential for the MW and a fit to RAVE and SDSS data \citep{Binney2012MNRAS.426.1328B}. However, he found  that  'the thick disc has to be hotter vertically than radially, a prediction that it will be possible to test in the near future'. However, we suspect that the stellar samples used to constrain his model were not large enough for the large number of free parameters of the model. To our knowledge, this prediction has not yet been confirmed and our model does not present this feature. In our model, the ratio $\sigma_R/\sigma_z$ for the thick disc is markedly smaller than that of the thin disc. 

In a much more recent study, \cite{2021A&A...647A..39S} attempted to adjust the so-called JJ Galaxy model \citep{2010MNRAS.402..461J} to \Gaia\  DR2 data using a MCMC scheme. These authors fitted density laws of the thin and thick discs together with their IMF and SFH. From the point of view of the kinematics, they consider an AVR following a power law as we do, and find results similar to ours for the thin disc. For the thick-disc vertical velocity dispersion, they find 44$\pm$4 \kms, which is in agreement with our value of 40.53 \kms. Interestingly, these authors model the thin disc with several star bursts with different vertical velocity dispersions, meaning a significantly larger number of free parameters than ours, at the expense of more flexibility. The main differences come from the data and observables used for their fit. They use the local sphere of 600 pc around the Sun from \Gaia\  DR2 and consider the W velocity as an observable, therefore relying on radial velocities, restricting the sample to bright stars and considering the kinematics only on the vertical axis. To complement those local data, \cite{2021A&A...647A..39S} consider the APOGEE DR14 sample of red clump stars to constrain the velocity space at larger distances from the Sun, with good radial velocities and distances. They further separate the thin- and thick-disc populations using the $\alpha$ abundances. However, their final sample of thin- and thick-disc populations from APOGEE is rather small (3910 and 847 respectively). Therefore, the extra information on the chemical separation of the two populations that they have is subject to Poisson noise in the MCMC fit. Finally, their approach is interesting from the point of view of the methodology and they will possibly extend it to larger samples and a wider range of Galactocentric radii in the future.

\cite{2019ApJ...878...21T} studied the vertical action as a function of Galactic radius and age for a sample of about 20\,000 red clump stars from the APOGEE DR14 spectroscopic survey, for which \Gaia\  DR2 proper motions are available. Distances were computed from Bayesian estimation and neural network computations were used to derive ages with a training sample coming from \Kepler\ asteroseismic parameters. \cite{2019ApJ...878...21T} fit a simple model to describe the evolution of the vertical action with age and \Rgal\ and conclude that it is possible to explain this evolution with only the scattering due to giant molecular clouds at \Rgal$<$10 kpc, while outward they estimate an initial vertical action at birth which grows fast with \Rgal, due to lower density and smaller vertical force. This allows the disc to flare quite significantly in the outer Galaxy. It is not easy to quantitatively compare our new model with theirs. However, qualitatively, our model does present a significant flare for the thin-disc components, but not for the thick-disc ones.

On the other hand, \cite{2019ApJ...878...21T} also made
N-body simulations in order to evaluate the gravitational evolution of disc galaxies. Among the many attempts, some simulated galaxies appear to have characteristics close to our MW, such that the processes at work in these simulations can be tentatively extrapolated to our own Galaxy. One of the studied processes is the evolution of the AVR.
A number of models have been developed in order to study the formation of the disc, and especially to explain the AVR by secular evolution of the disc due to the effect of giant molecular clouds, spiral waves, or even bar perturbations, or by heating due to mergers \citep{Dehnen1998MNRAS.298..387D,Aumer2016MNRAS.462.1697A}. In particular, \cite{Aumer2016MNRAS.462.1697A} analysed N-body simulations and explained the differences between the observed AVR and the heating history. These authors assess that the observed AVR suffers from age uncertainties, which tends to flatten the relations (lowering the value of $\beta$). Their diverse simulations underwent very different formation histories leading to different AVRs, showing a variety of possibilities from the point of view of Galaxy evolution, depending on the giant molecular clouds (GMC) masses, bar strength, and spiral waves. \cite{Aumer2016MNRAS.462.1697A} also claimed that there is no universal heating history valid for all populations, because the heating process significantly evolves over time and is different from one Galactic region to another. The analysis 
of chemodynamical simulations shed light on the formation scenarios of various populations, helping to identify the efficient processes at work, but it is difficult to link a variety of simulations with different histories with the Milky Way evolution. In our complementary approach, we ensure a realistic potential is obtained at the present time, which is interesting to compare with various simulations. In the case of the AVR, our result is compatible with a secular heating of the thin disc by 
GMC, but we cannot differentiate this process from other efficient heating processes.
 On the other hand, in \cite{2018MNRAS.476.3648N}, hydrodynamic simulations of a Galaxy similar to the Milky Way exhibit an AVR similar to the one in our model, coming from the settling of the gas only, and not from secular heating and radial migration. This new scenario is an interesting alternative to the secular heating by GMCs or spiral waves, although it is difficult to estimate how close this simulation is to the real MW formation scenario.

With its various characteristics (compactness, dropping density in the outer Galaxy, thickness, large distinction in velocity dispersion ellipsoid with the thin disc, and no flare), we favour the scenario where the thick disc is formed from a turbulent gas disc in the early Galaxy, as proposed by \cite{Bournaud2009}.

Several chemodynamical simulations have been specifically developed to study the Milky Way formation and to compare with distributions of velocities and abundances. Among them, \cite{2013A&A...558A...9M} and \cite{2014A&A...572A..92M} used N-body simulations from \cite{2012ApJ...756...26M}, adding a chemical evolution scenario to `paint' their particles and evaluate the chemo-kinematical distributions to be compared with real data. This allowed them to evaluate stellar migration and the impact of mergers, and helped them to interpret spectroscopic surveys in terms of the Galactic formation scenario. Among other interesting results, \cite{2017ApJ...834...27M} found (their figure 7) that, in their simulations, all low-$\alpha$ mono-age populations flare \citep{2015ApJ...804L...9M} while high-$\alpha$ populations do not, completely in line with our result. 

\subsection{Rotation curve and the Galaxy mass}

Recent \Gaia\  observations allowed a precise determination of the shape of the Galactic rotation curve, mainly through measurements of Cepheids within a few kiloparsecs of the Sun and at greater distances up to 60 kpc from the Galactic centre using the kinematics of halo stars \citep{2020MNRAS.494.4291C,2021MNRAS.501.5964D}. Evidence of a rapid decay of the Galactic rotation curve has led to a revision of the mass of the Galaxy included in a radius of 200 kpc. We adopted these measurements as constraints on our model, and therefore we have by construction the same total mass as the latter authors: 1.08$^{+0.20}_{-0.14} 10^{12}$ \Msun\ at 200 kpc \citep{2020MNRAS.494.4291C} in good agreement with the recent measurements of \cite{2019MNRAS.484.5453C} who found M$_{200}=1.17^{+0.21}_{-0.15}\times 10^{12}$\Msun\ based on the dynamics of satellites. \cite{2021MNRAS.501.5964D} determined the Milky Way mass before and after the infall of the 
Large Magellanic Cloud (LMC). These authors found a total mass at 200 kpc of 1.01 $\pm 0.24 \times 10^{12}$ \Msun\ before, or 1.16 $\pm 0.24 \times 10^{12}$ \Msun\ after.

Recently, \cite{2021MNRAS.501.2279V} measured the influence of the LMC passage on the Galactic mass determination. The influence of these external perturbations remains relatively minor for the inner parts  \citep[Figure 15 in  ][]{2021MNRAS.501.2279V} and in particular the definition of a mean rotation curve remains meaningful.

\section{Conclusions and perspectives}
\label{Sect:conclusion}

We obtained a self-consistent dynamical model in good agreement with astrometric eDR3 \Gaia\  data. The axisymmetric numerical gravitational potential presented here  produces both density laws and kinematics that are consistent with these data, locally and at larger distances from the Sun, in the range of 6 to 12 kpc in Galactocentric radius and $-2$ to 2 kpc in $z$, although the model can be extrapolated much further.

        This model combines several elements that are rarely found together in other Galactic models. The dynamical coherence links the gravitational potential to all the Galactic components. The stellar distribution functions, density and kinematics, are themselves consistent with the potential. 
   
        The approach  involves less uncertainty than previous ones for the following reasons: (i)   we provide dynamically justified stellar vertical and horizontal density distributions, (ii)  we make use of a stellar luminosity function constrained by stellar mass luminosity relationships provided by the most recent evolutionary tracks and also constrained by recent \Gaia\  observations; and (iii) fitting is done simultaneously on density and kinematics in the space of observables. {\it Starevol} stellar evolutionnary models  \citep[e.g.][]{2012A&A...543A.108L,2019A&A...631A..77A} are state-of-the-art interior models covering a wide mass range from 0.6 to 6 solar masses, and similarly wide ranges of metallicity and $\alpha$-abundance, which have been incorporated into the BGM by \cite{2017A&A...601A..27L,2019A&A...621A..24L}. They ensure the reliability of the mass--luminosity relation in this mass range.

Our model presents relevant contributions to the modelling of the Milky Way, in particular :
\begin{itemize}
    \item We find the solar motion with respect to the LSR to be \usun=10.79 $\pm$ 0.56 \kms, \vsun=11.06 $\pm$ 0.94 \kms, and \wsun=7.66 $\pm$ 0.43 \kms, in good agreement with recent studies.
    \item The thin disc is modelled by the sum of seven age components, each having a specific velocity ellipsoid. We determined its AVR at different positions in the Galaxy and showed that velocity dispersions  vary with $R$ and $z$ and the kinematical scale lengths are about three times the density scale lengths.
    \item In the range of Galactic radii 6-12 kpc, the $\sigma_R/\sigma_z$ varies with age, $R,$ and $z$ and has values of between 1.9 and 2.6 in the thin discs (for $z<2$ kpc) and between 1.2 and 1.4 in the thick disc.
    \item The asymmetric drift is increasing with age and $z$ and decreasing with $R$. It can reach high values (more than 80 \kms) at high $z$ in the thick disc. This lag is directly obtained from the potential and the self-consistent distribution functions.
    \item The tilt of the velocity ellipsoid is also varying with age, $R,$ and $z,$ but the variations are small and it nearly follows spherical symmetry (major axis nearly pointing towards the Galactic center) for all populations.
    \item The thin disc population density exhibits a significant flare in the outskirts of the Milky Way, coming naturally from the variation of the vertical force with $R$. On the contrary, the thick disc does not flare, probably because of its compactness and the fast drop in density in the outer Galaxy. 
    \item The young thick-disc characteristics are significantly different from the old thin-disc populations, pointing towards a clear distinction between them. In particular, we obtain an asymmetric drift in the thick disc that is about twice that in the old thin disc
 at the Solar position and very different variations with $R$ for the mean $V_\phi$, and for $\sigma_\phi$. This result reinforces the idea of a very different scenario of formation between the thin and thick discs.
\end{itemize}

We have not considered the SFH of the disc as a free parameter. It was determined using the analysis of \cite{2019A&A...624L...1M} using \Gaia\  DR2 up to apparent magnitude $G=13$. Were this SFH inadequate for our study, it could slightly impact the present analysis. However, if this were the case, the self-consistent model would be revised applying again the present method with corrected SFH in the initial simulations. 

On the low-mass side of the IMF, the stellar mass is subject to uncertainties because stellar models are less accurate. Their atmospheres are difficult to model because of the numerous molecules.
We did not consider them in our analysis for those reasons. However, their number constitutes a source of uncertainty when considering the total mass in stars in the Milky Way.
This will be considered in a future study where the IMF and SFH will be explored extensively and the impact on the dynamical model will be estimated.

We find that, despite the very good agreement between the axisymmetric model and the data, there are significant discrepancies in a few fields where further studies need to be conducted. This is true particularly towards the anticentre, where disc substructures have been highlighted, and towards the north Galactic pole where a vertical wave is seen. We also point that our stellar halo model needs to be improved, but as this population is very minor in mass density, this does not preclude the use of the present Galactic potential. 

The model can further be used as a realistic potential of the Milky Way ---in orbital studies for example--- to derive eccentricities, energy, and angular momentum of real stars from accurate astrometric data such as \Gaia\  and other space surveys to come.
We believe that it can also be used for identifying and characterising substructures, allowing the user to subtract the axisymmetric component from the data to enhance the contrast and to qualify the interesting features.
Furthermore, we expect that the recently released \Gaia\  DR3 and large spectroscopic surveys will provide new opportunities to explore the gravitational potential of our Galaxy  in even greater detail.

\begin{acknowledgements}

We are very pleased to pay tribute to Michel Cr\'ez\'e who instigated and initiated this work in the early 1980s. This work made use of the Topcat \url{ http://www.starlink.ac.uk/topcat/} and stilts \url{ http://www.starlink.ac.uk/stilts/} softwares and are very glad to thank Mark Taylor for doing continuous improvements on these tools.

We acknowledge the financial support from "Programme National Cosmologie et Galaxies" (PNCG) and "Programme National de Physique Stellaire" (PNPS) of CNRS/INSU, France.
BGM simulations and MCMC fits were executed on computers from the Utinam Institute of the
Universit\'e de Franche-Comt\'e, supported by the R\'egion de
Franche-Comt\'e and Institut des Sciences de l'Univers (INSU).

This work has made use of data from the European Space Agency (ESA) mission
{\it Gaia} (\url{https://www.cosmos.esa.int/gaia}), processed by the {\it Gaia}
Data Processing and Analysis Consortium (DPAC,
\url{https://www.cosmos.esa.int/web/gaia/dpac/consortium}). Funding for the DPAC
has been provided by national institutions, in particular the institutions
participating in the {\it Gaia} Multilateral Agreement. 

We also acknowledge the International Space Science Institute, Bern, Switzerland for providing financial support and meeting facilities.

J.G.F-T gratefully acknowledges the grant support provided by Proyecto Fondecyt Iniciaci\'on No. 11220340, and also from ANID Concurso de Fomento a la Vinculaci\'on Internacional para Instituciones de Investigaci\'on Regionales (Modalidad corta duraci\'on) Proyecto No. FOVI210020, and from the Joint Committee ESO-Government of Chile 2021 (ORP 023/2021). 
\end{acknowledgements}

\bibliographystyle{aa} 
\bibliography{dyco} 

\begin{appendix}
\section{MCMC variables dependencies}
\label{App:correlations}

\begin{figure*}[h!]
\label{fig:correlation}
\centering
\begin{tabular}{lllllll}
\subf{\includegraphics[width=0.12\textwidth]{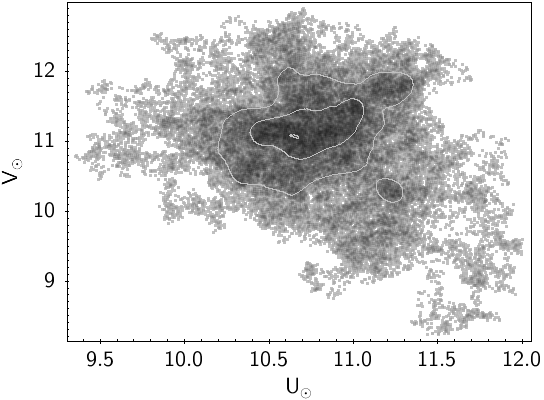}}{} & \\
\subf{\includegraphics[width=0.12\textwidth]{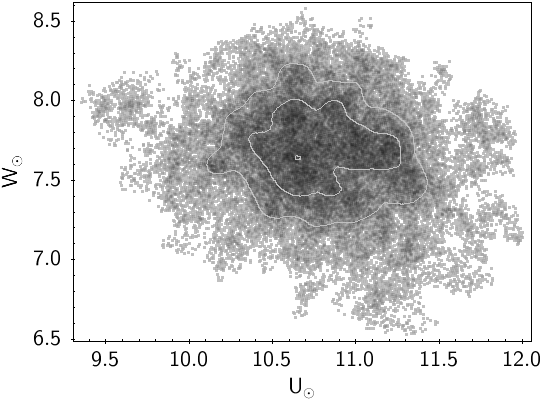}}{} &
\subf{\includegraphics[width=0.12\textwidth]{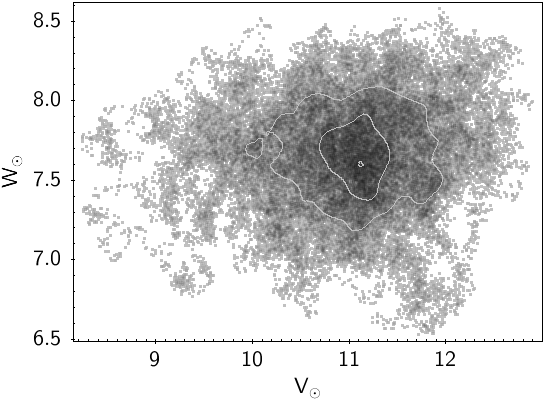}}{} & & & & & \\
\subf{\includegraphics[width=0.12\textwidth]{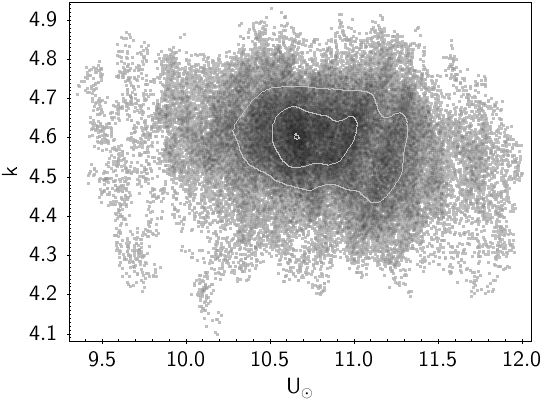}}{} &
\subf{\includegraphics[width=0.12\textwidth]{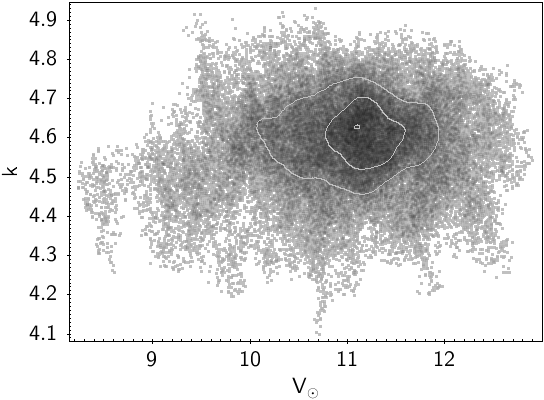}}{} &
\subf{\includegraphics[width=0.12\textwidth]{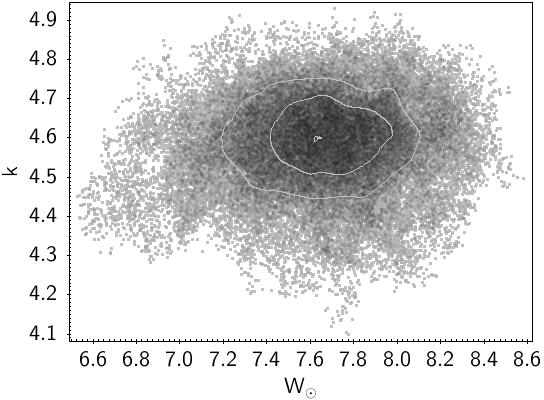}} {}& & & &\\
\subf{\includegraphics[width=0.12\textwidth]{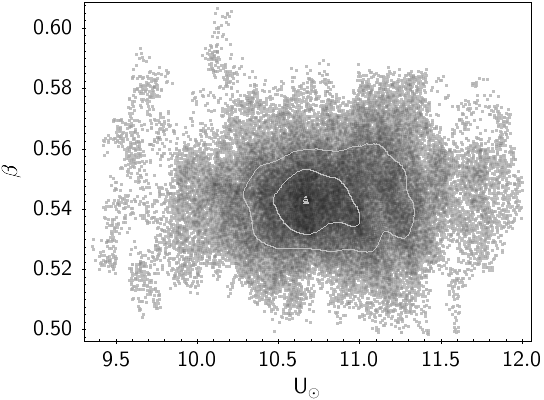}}{} & 
\subf{\includegraphics[width=0.12\textwidth]{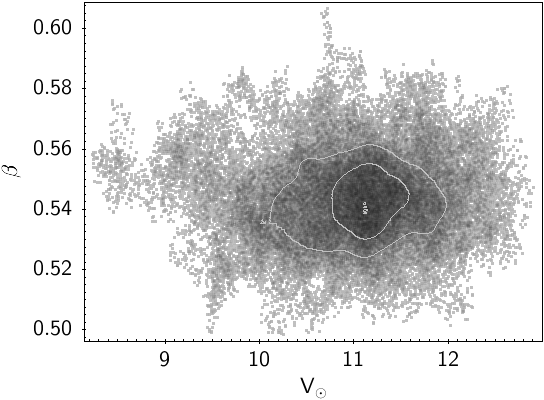}}{}  &
\subf{\includegraphics[width=0.12\textwidth]{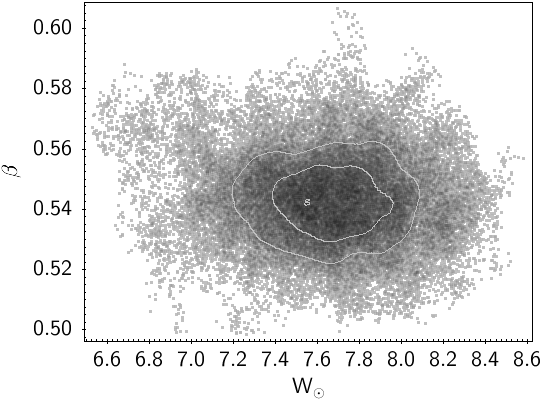}}{} &
\subf{\includegraphics[width=0.12\textwidth]{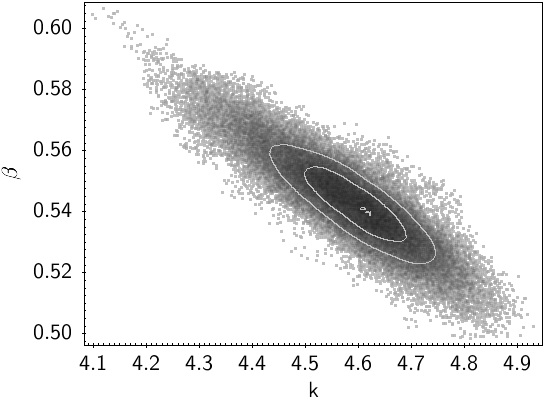}}{} & & &\\
\subf{\includegraphics[width=0.12\textwidth]{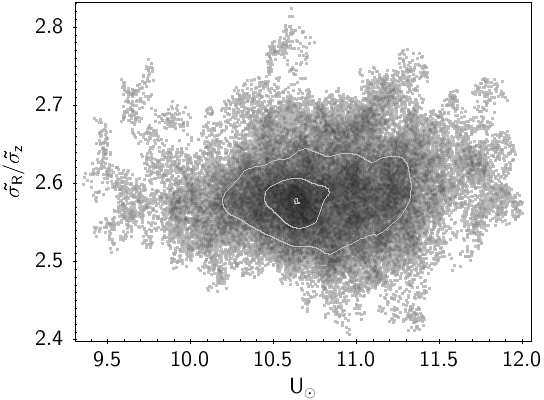}}{} &
\subf{\includegraphics[width=0.12\textwidth]{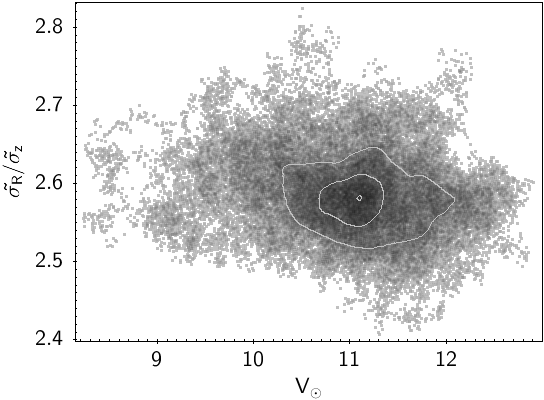}}{} &
\subf{\includegraphics[width=0.12\textwidth]{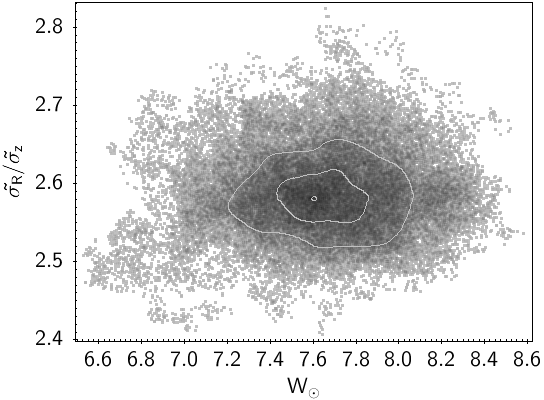}}{}  & 
\subf{\includegraphics[width=0.12\textwidth]{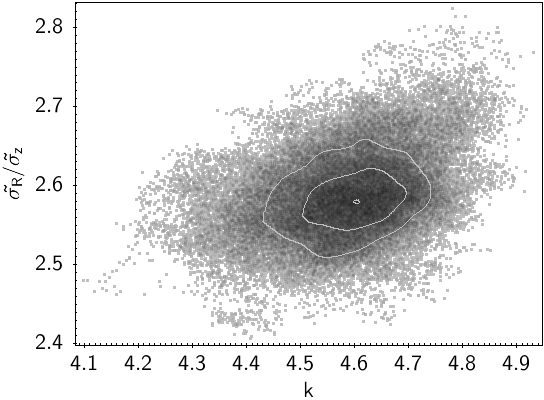}} {} &
\subf{\includegraphics[width=0.12\textwidth]{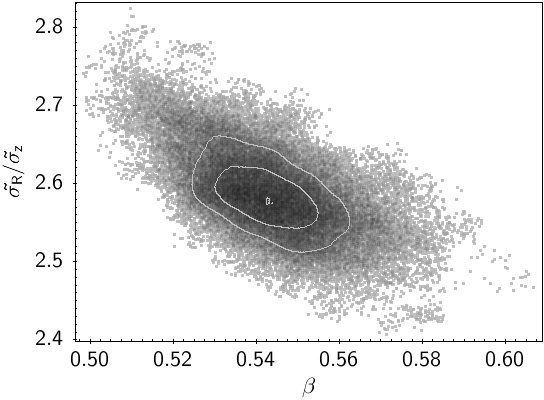}}{} & & \\
\subf{\includegraphics[width=0.12\textwidth]{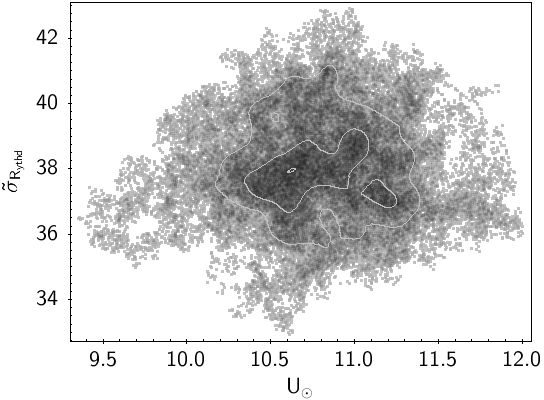}}{} &
\subf{\includegraphics[width=0.12\textwidth]{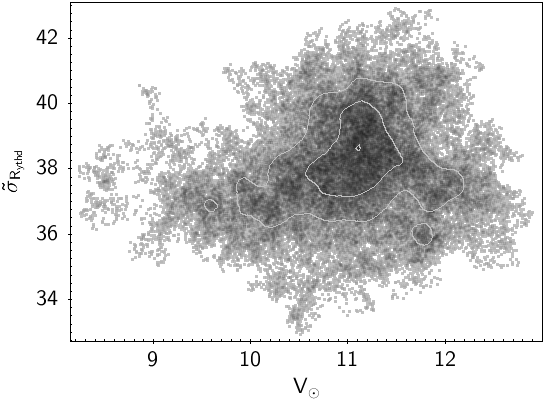}}{} &
\subf{\includegraphics[width=0.12\textwidth]{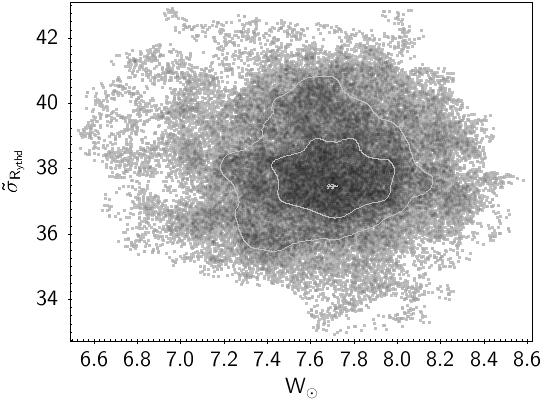}}{} &
\subf{\includegraphics[width=0.12\textwidth]{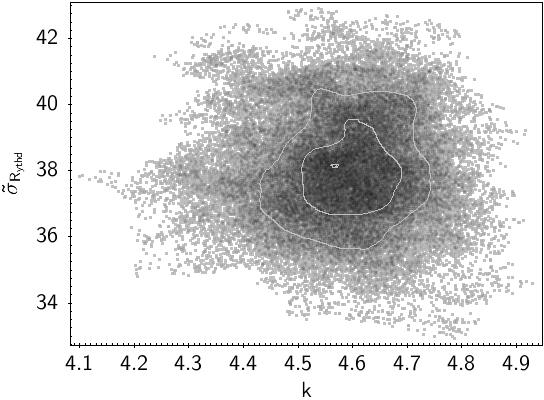}}{}  &
\subf{\includegraphics[width=0.12\textwidth]{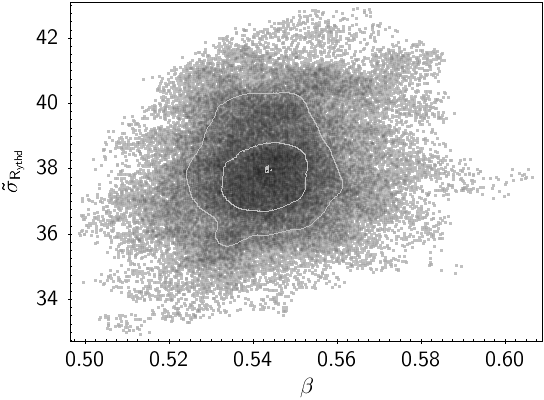}}{} & 
\subf{\includegraphics[width=0.12\textwidth]{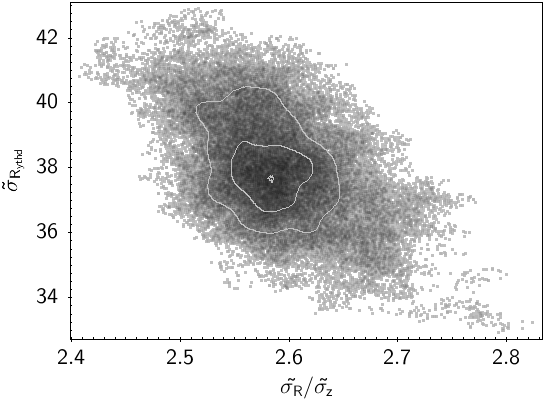}}{} & \\
\subf{\includegraphics[width=0.12\textwidth]{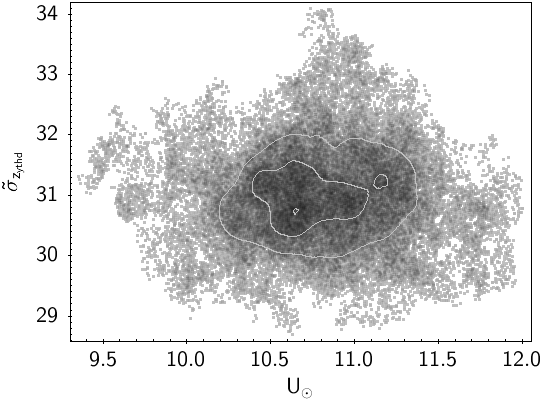}}{} &
\subf{\includegraphics[width=0.12\textwidth]{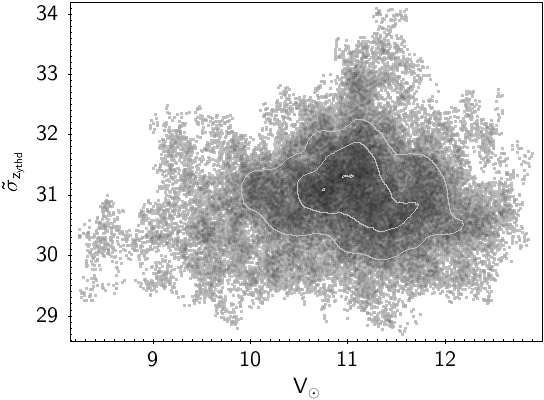}}{} &
\subf{\includegraphics[width=0.12\textwidth]{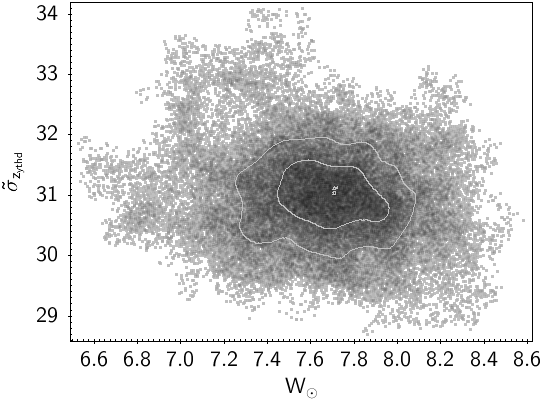}}{} &
\subf{\includegraphics[width=0.12\textwidth]{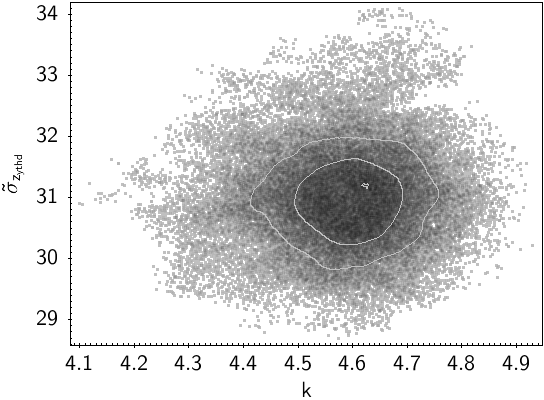}} {} &
\subf{\includegraphics[width=0.12\textwidth]{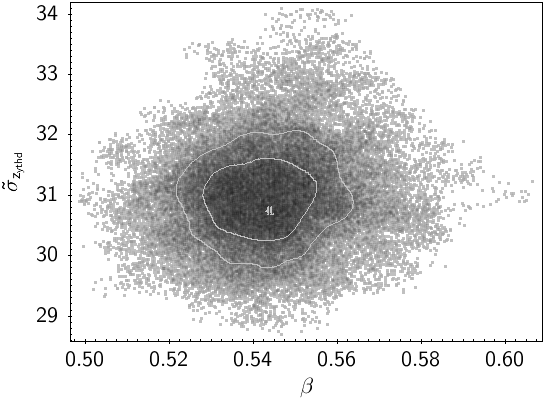}} {} &
\subf{\includegraphics[width=0.12\textwidth]{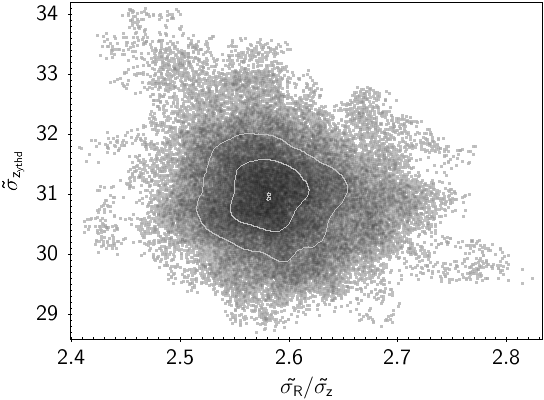}} {} &
\subf{\includegraphics[width=0.12\textwidth]{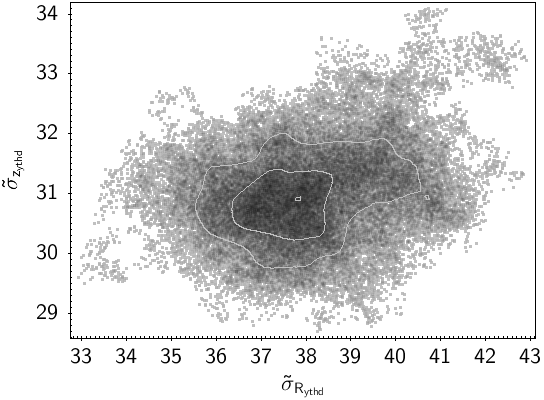}} {} \\
\end{tabular}
\caption{Triangular plot of the parameter values of the best MCMC chains.}
\end{figure*}

We present in Fig.\ \ref{fig:correlation} a triangular plot showing the correlations between kinematic parameters of the best Markov Chains. A significant anti-correlation is found between the AVR parameters $k$ and $\beta$ but this does not imply a bad determination of the AVR itself because the corresponding shapes of the AVR are very similar. There is also a significant anti-correlation between the thin disc \sigratio\ and the \sigR of thick disc due to the fact that both populations are mixed up in a wide proportion of the $(R,z)$ space. The third correlation is between thin disc \sigratio\ and $\beta$ due to the fact that we have only two components of the velocity and we miss the line-of-sight velocity.

\section{Density and velocity distributions of the final model}
\label{App:Rz}

\begin{landscape}
\begin{figure}[h!]
\begin{center}
\includegraphics[width=3cm]{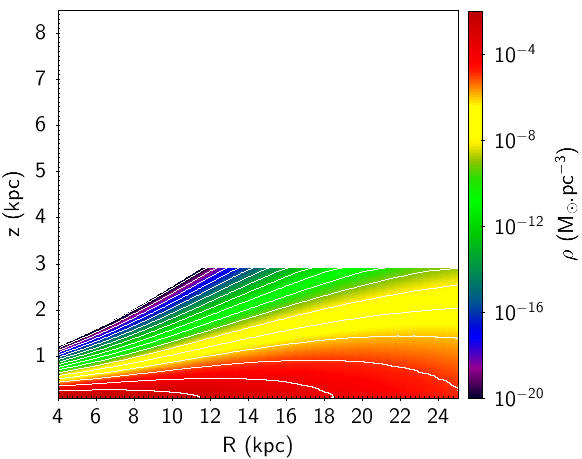}
\includegraphics[width=3cm]{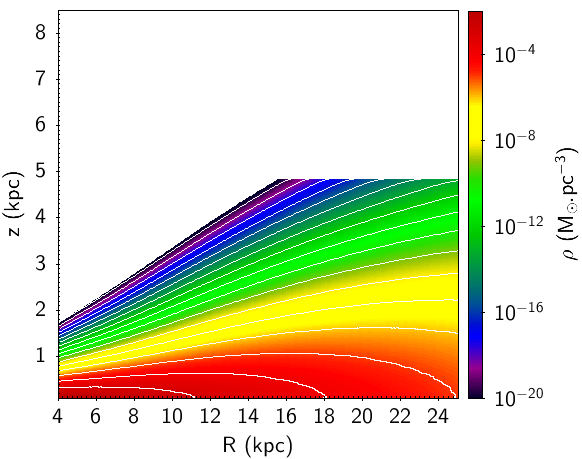}
\includegraphics[width=3cm]{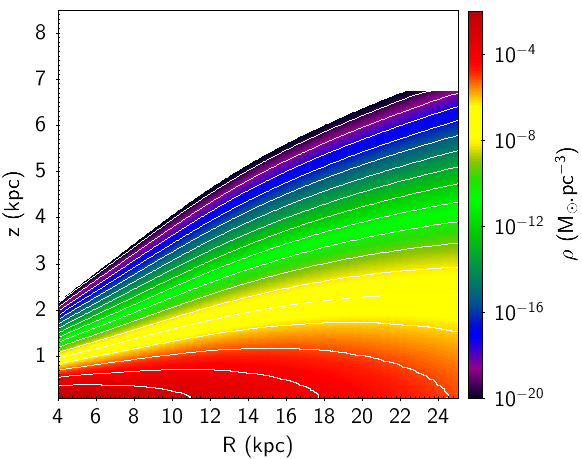}
\includegraphics[width=3cm]{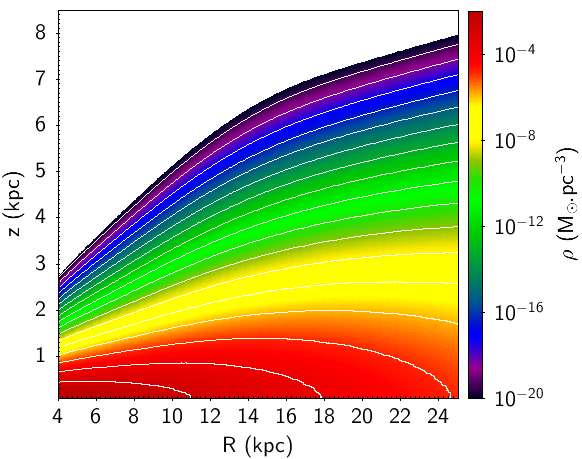}
\includegraphics[width=3cm]{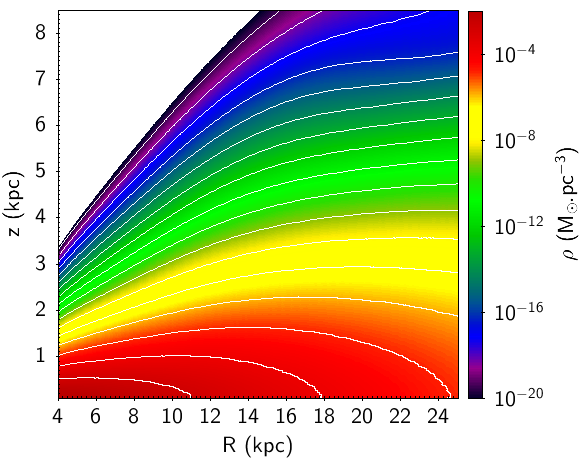}
\includegraphics[width=3cm]{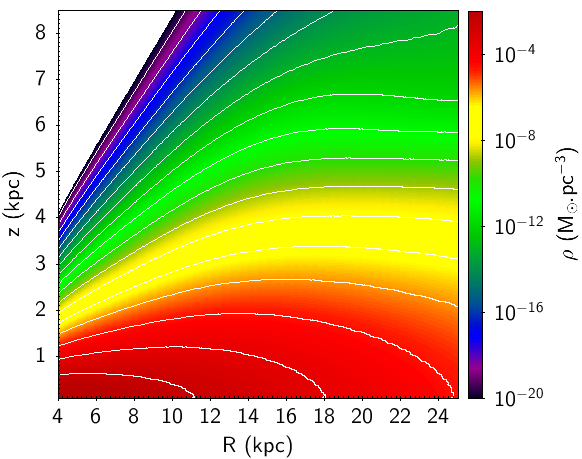}
\includegraphics[width=3cm]{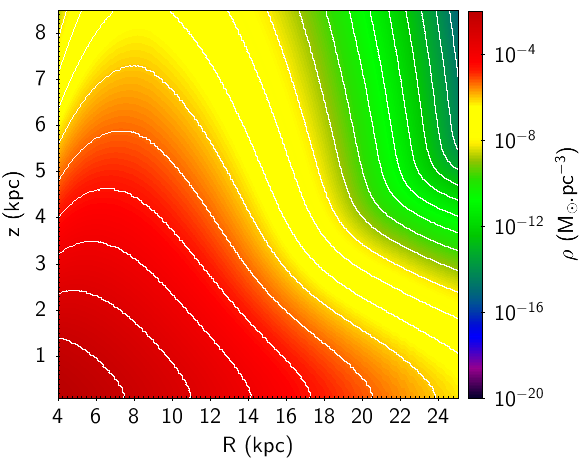}

\includegraphics[width=3cm]{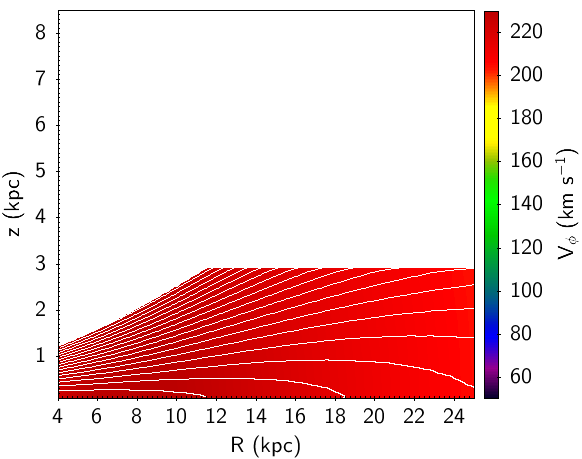}
\includegraphics[width=3cm]{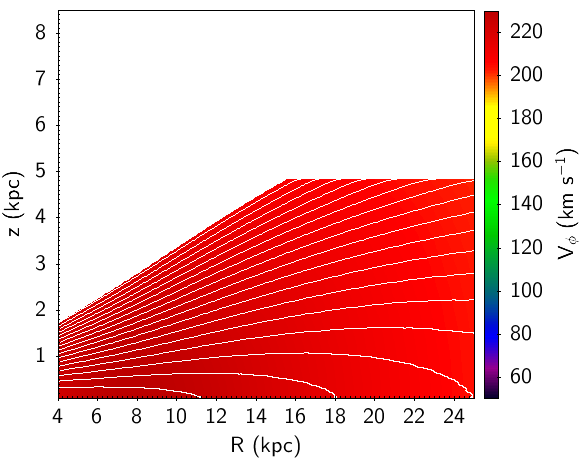}
\includegraphics[width=3cm]{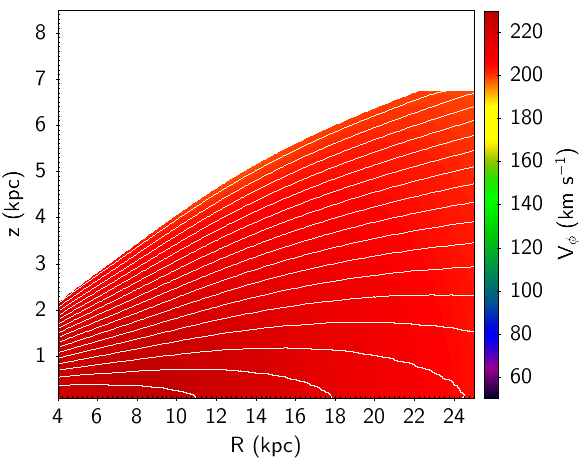}
\includegraphics[width=3cm]{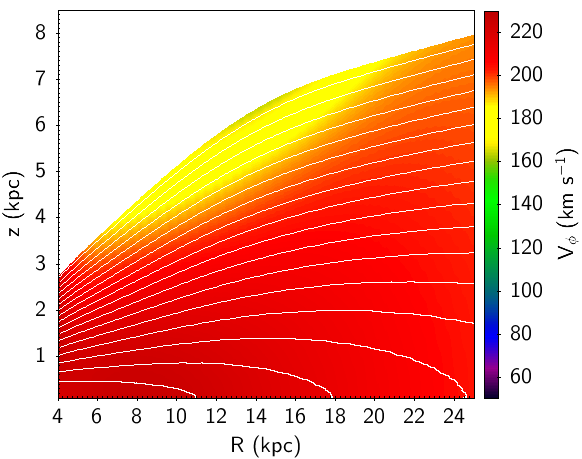}
\includegraphics[width=3cm]{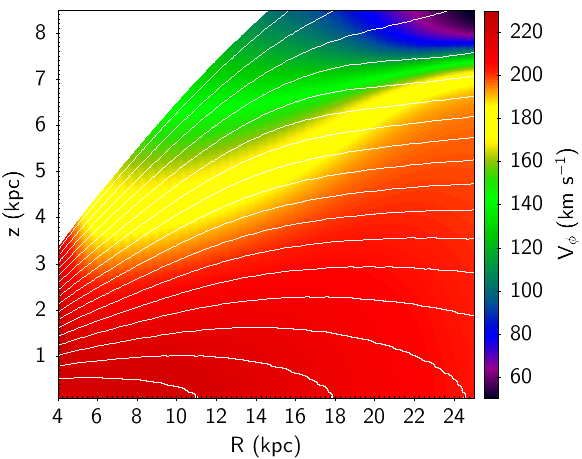}
\includegraphics[width=3cm]{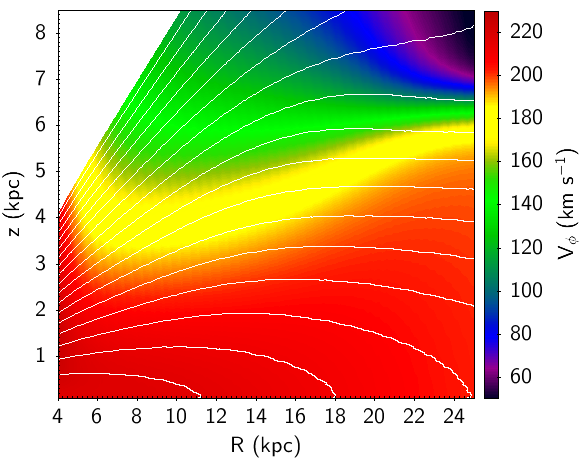}
\includegraphics[width=3cm]{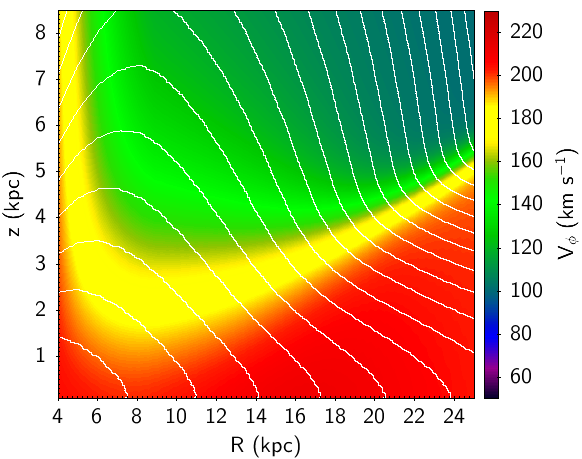}

\includegraphics[width=3cm]{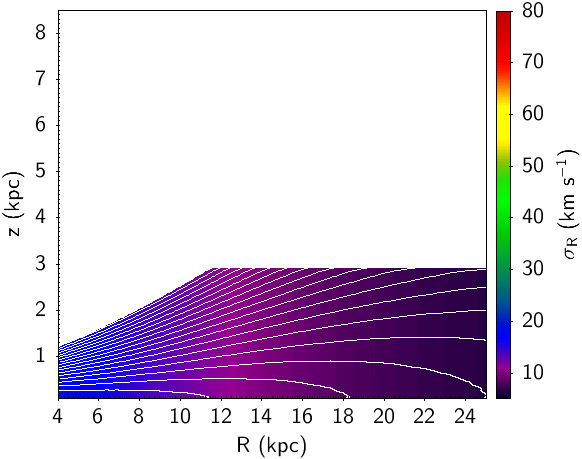}
\includegraphics[width=3cm]{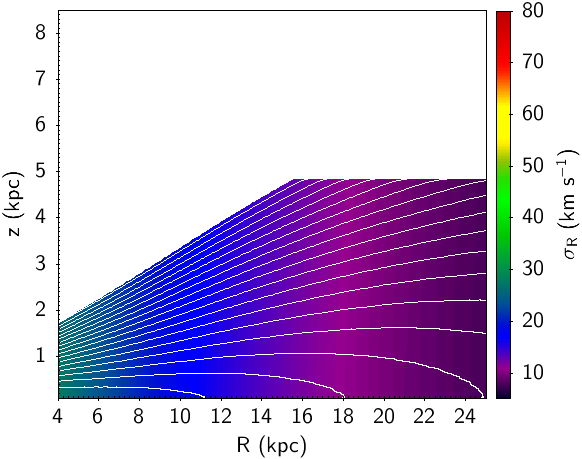}
\includegraphics[width=3cm]{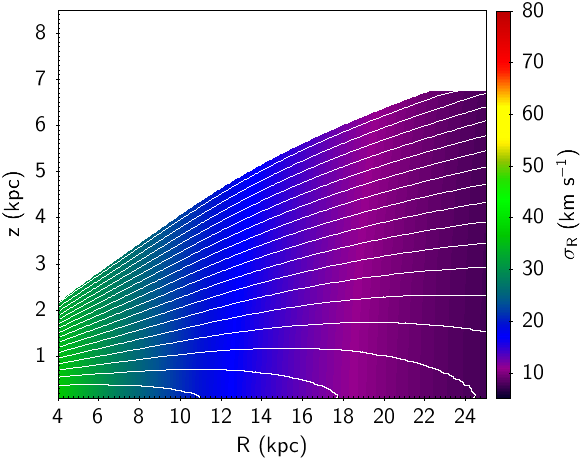}
\includegraphics[width=3cm]{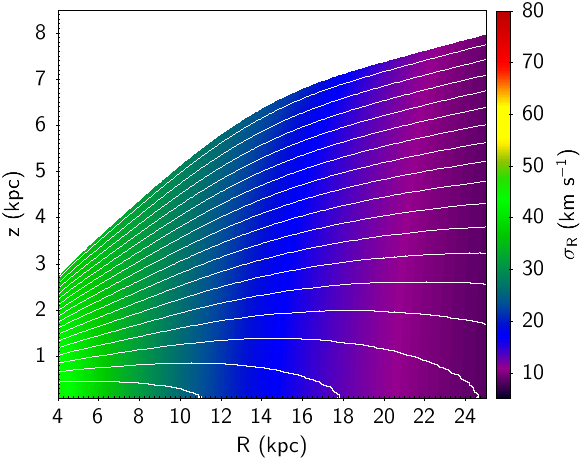}
\includegraphics[width=3cm]{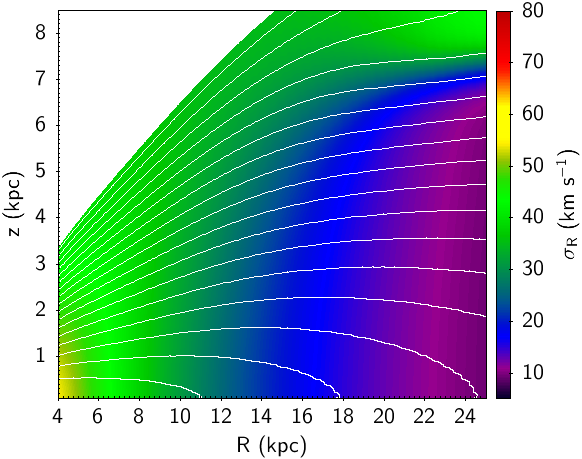}
\includegraphics[width=3cm]{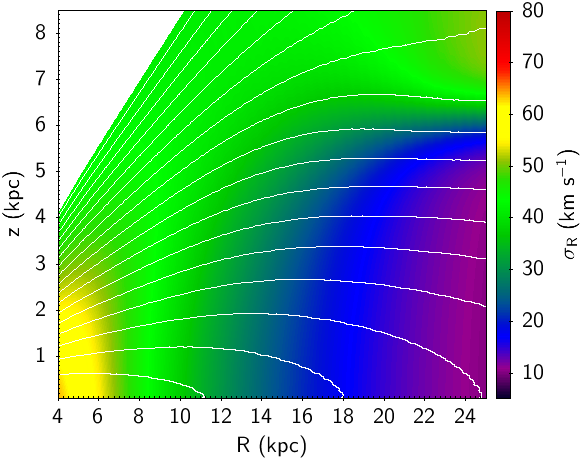}
\includegraphics[width=3cm]{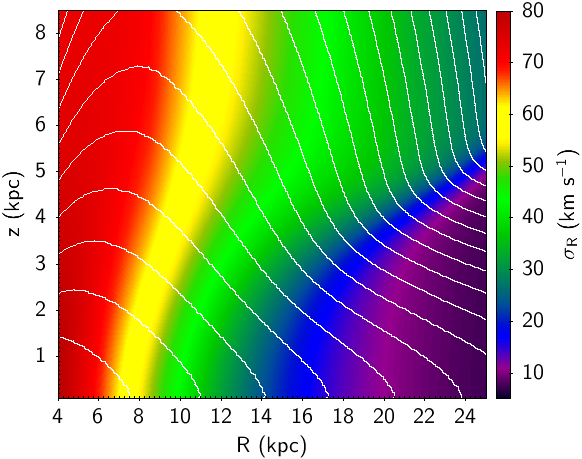}

\includegraphics[width=3cm]{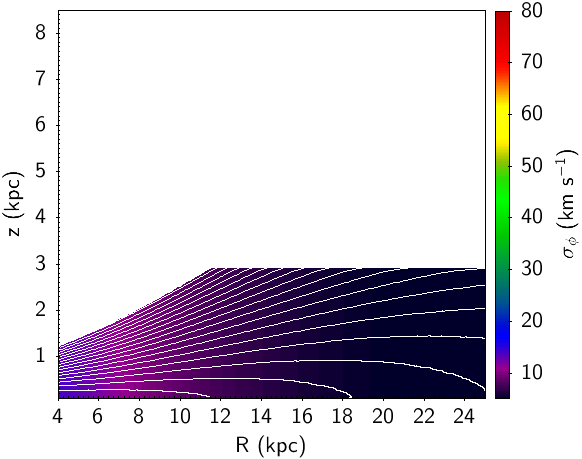}
\includegraphics[width=3cm]{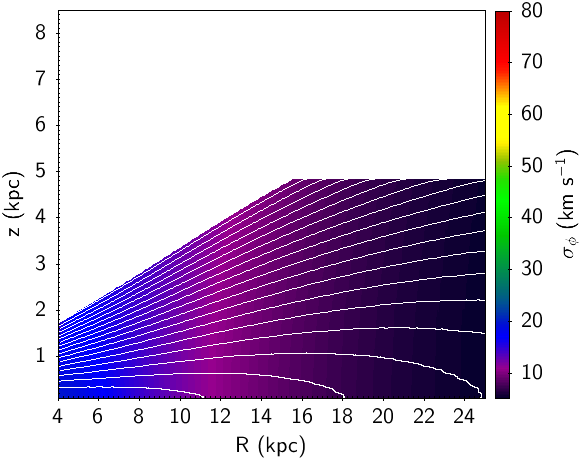}
\includegraphics[width=3cm]{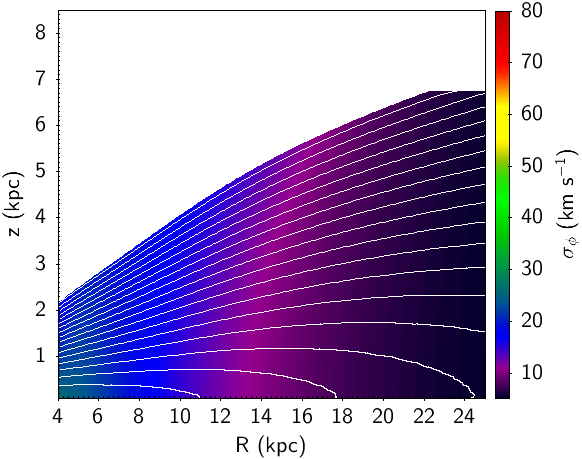}
\includegraphics[width=3cm]{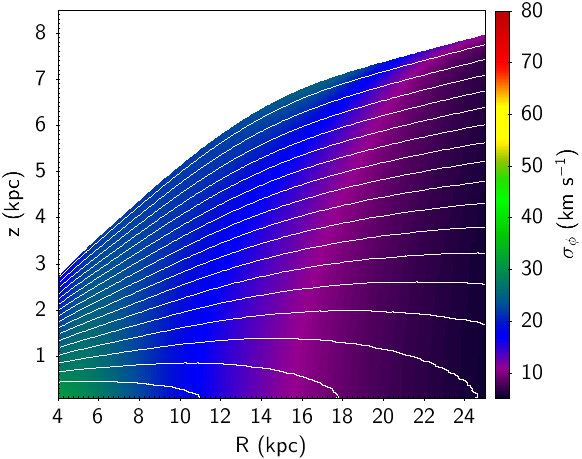}
\includegraphics[width=3cm]{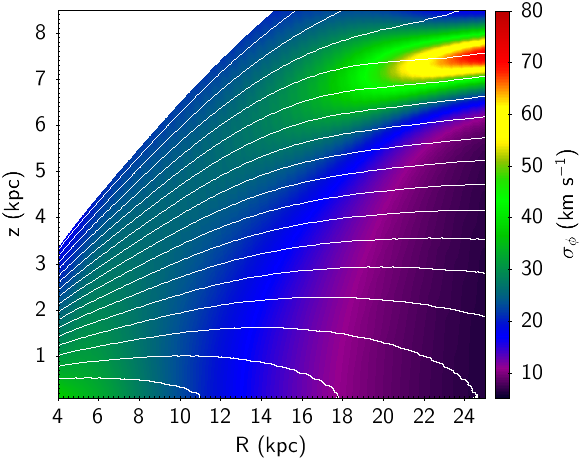}
\includegraphics[width=3cm]{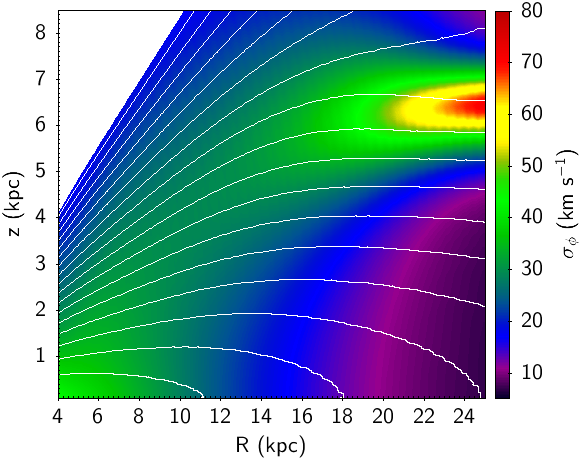}
\includegraphics[width=3cm]{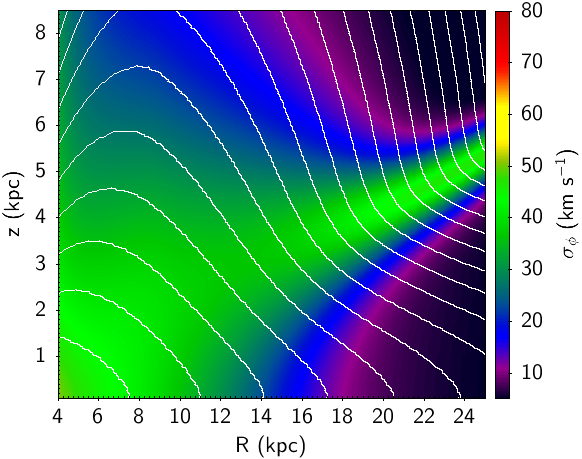}

\includegraphics[width=3cm]{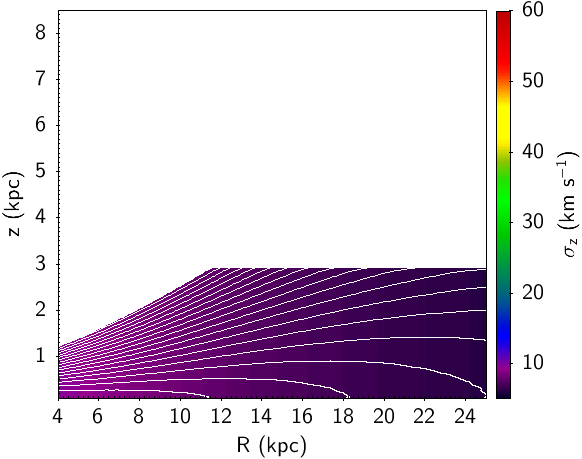}
\includegraphics[width=3cm]{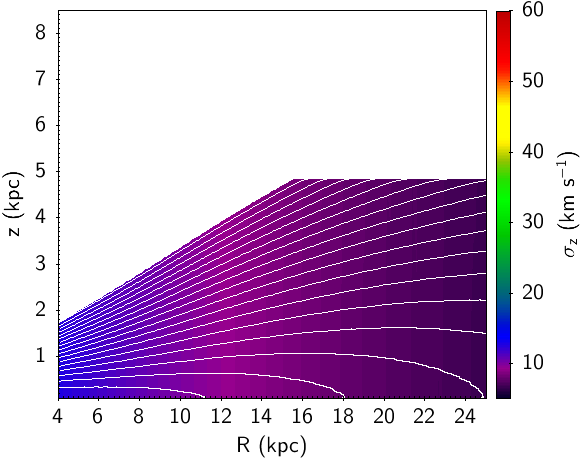}
\includegraphics[width=3cm]{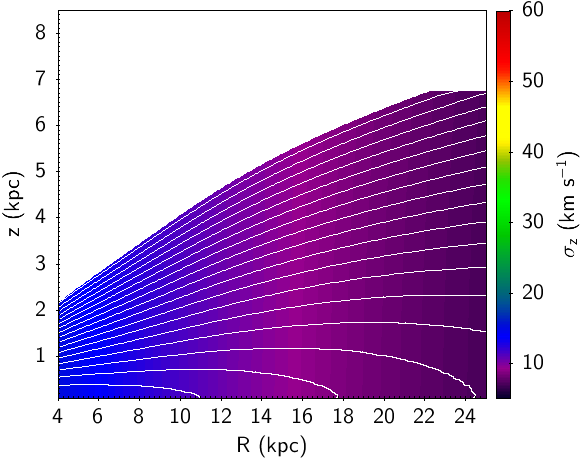}
\includegraphics[width=3cm]{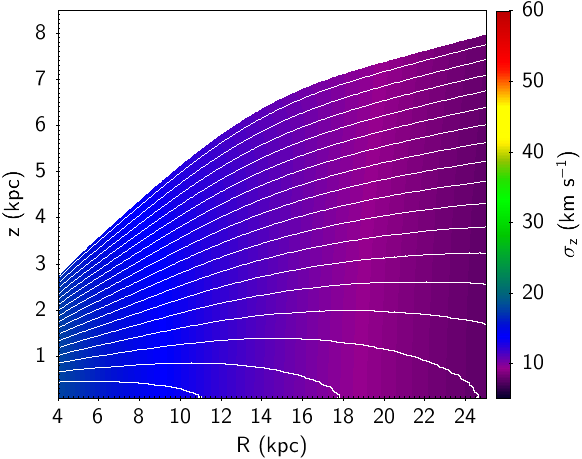}
\includegraphics[width=3cm]{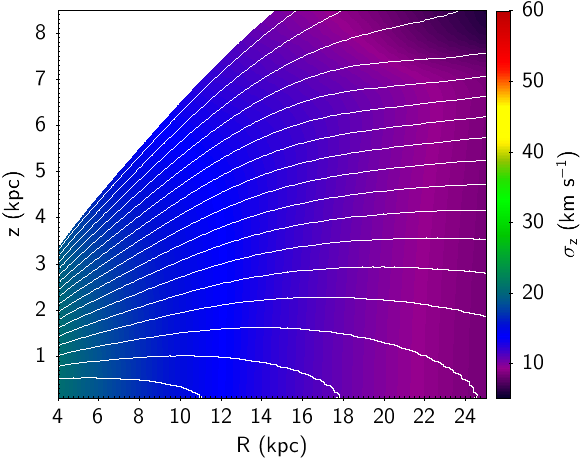}
\includegraphics[width=3cm]{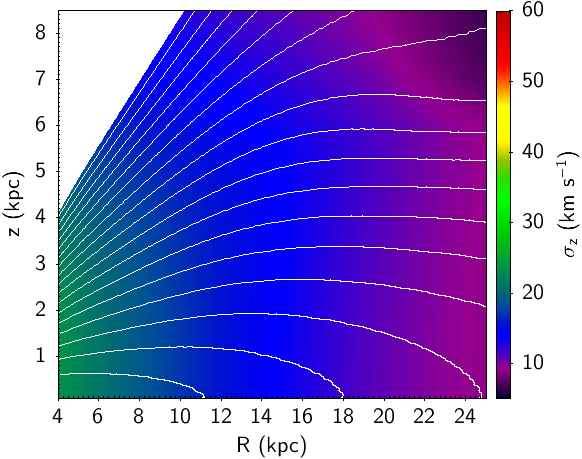}
\includegraphics[width=3cm]{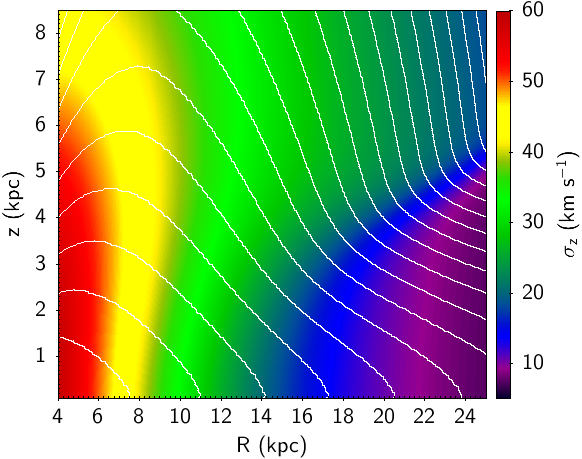}

\includegraphics[width=3cm]{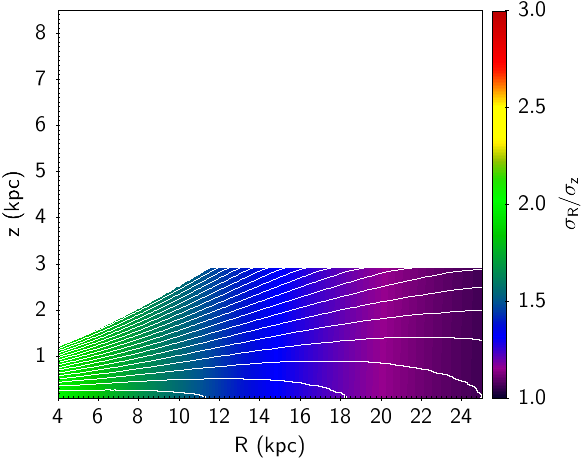}
\includegraphics[width=3cm]{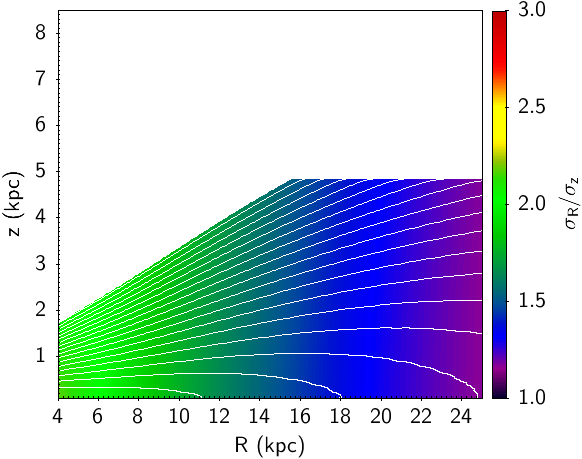}
\includegraphics[width=3cm]{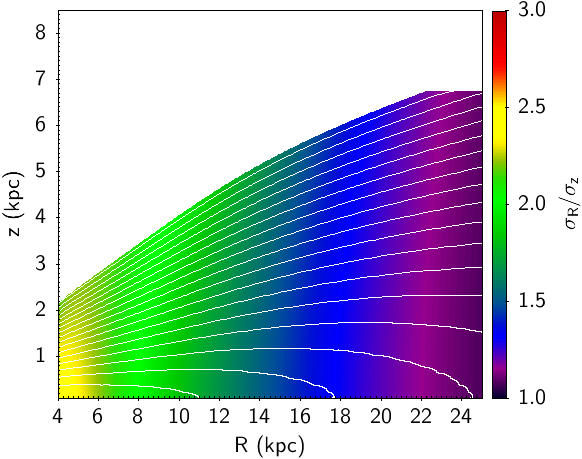}
\includegraphics[width=3cm]{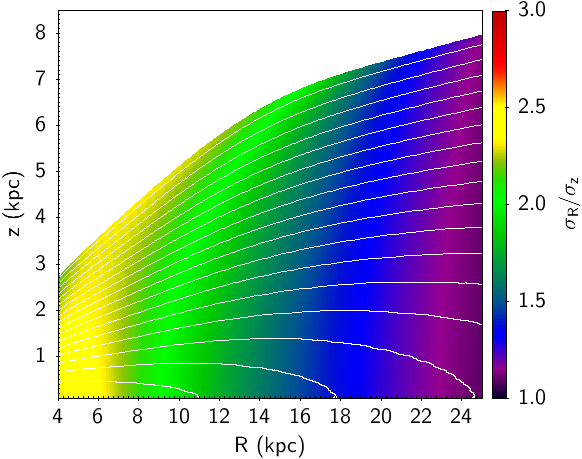}
\includegraphics[width=3cm]{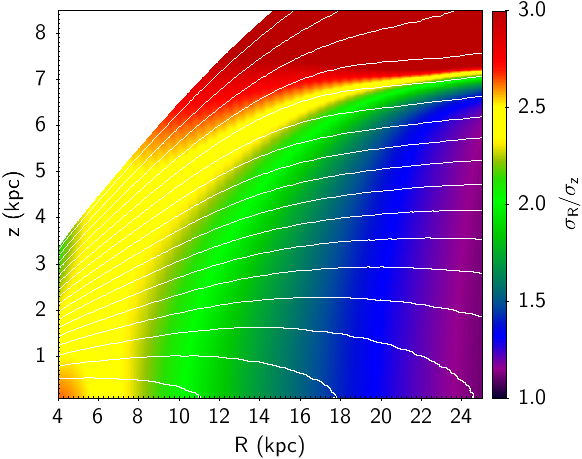}
\includegraphics[width=3cm]{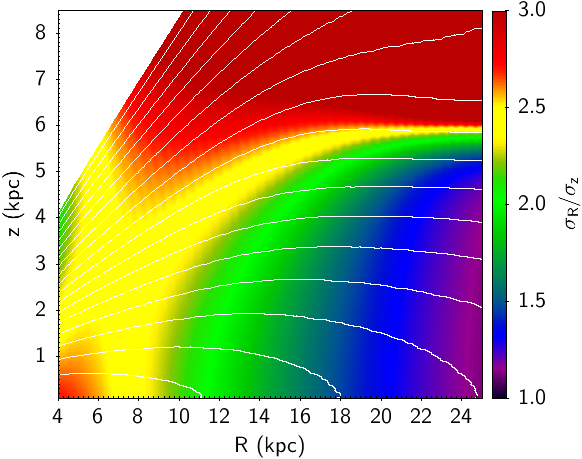}
\includegraphics[width=3cm]{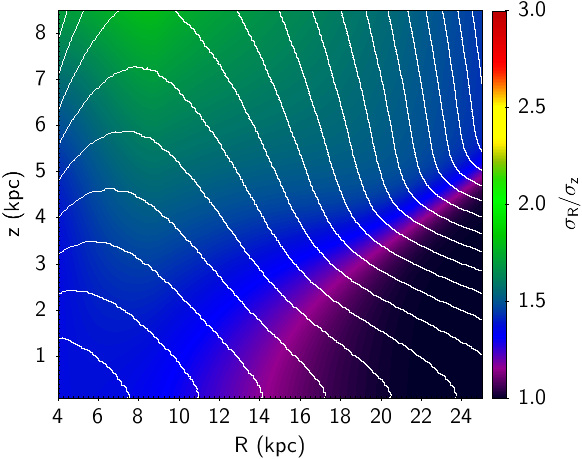}

\includegraphics[width=3cm]{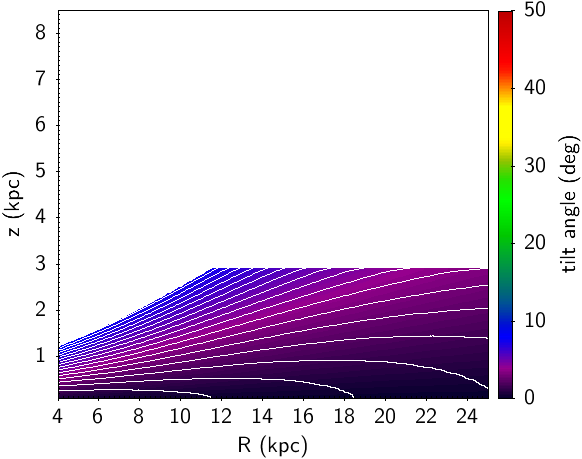}
\includegraphics[width=3cm]{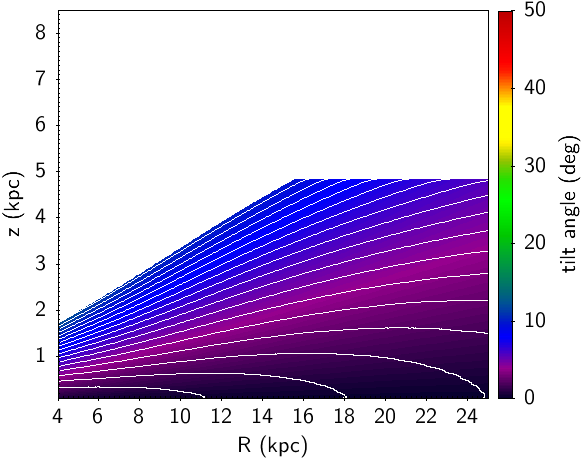}
\includegraphics[width=3cm]{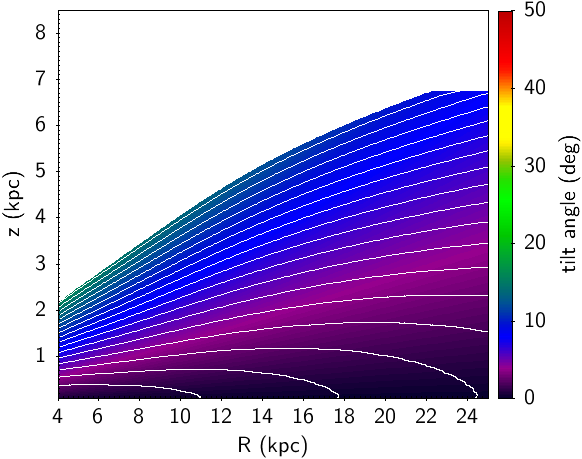}
\includegraphics[width=3cm]{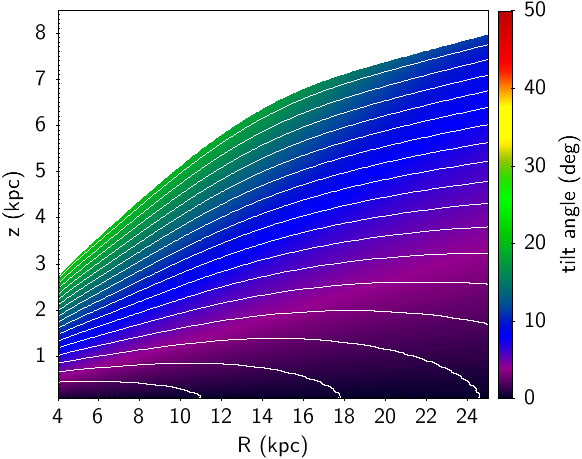}
\includegraphics[width=3cm]{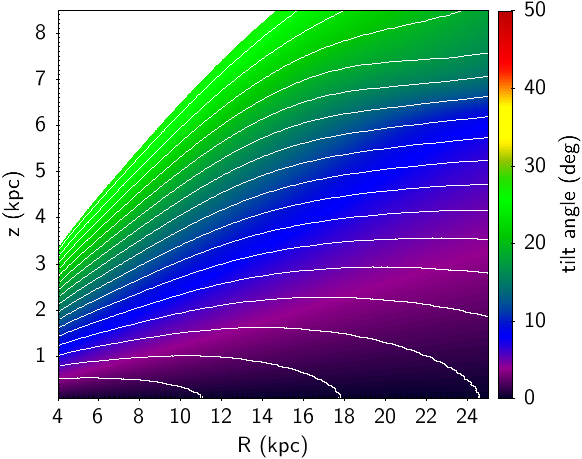}
\includegraphics[width=3cm]{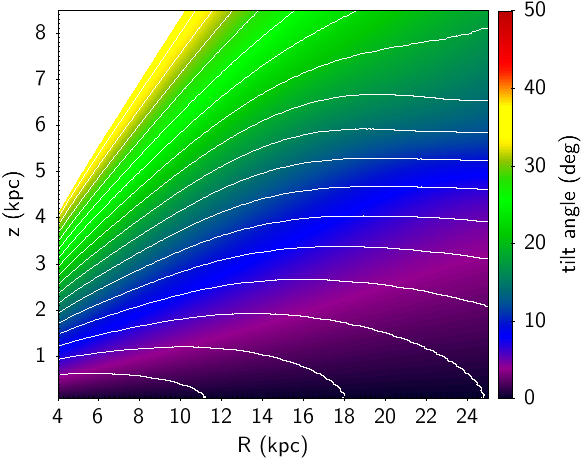}
\includegraphics[width=3cm]{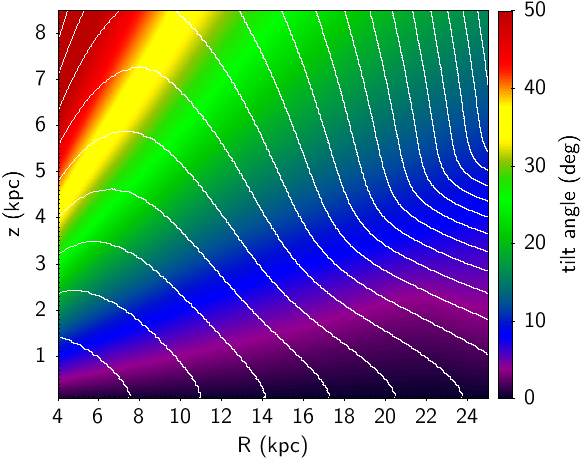}

\caption{Density and kinematics as a function of $R$ and $z$ for different disc components. Columns 1 to 6 represent thin disc components 2 to 7. Columns 7 is for the young thick disc.  
The densities are shown for values larger than $10^{-20}\ M_{\odot}\ pc^{-3}$. Regions in white have either not been computed or the density is below this limit. Different rows are for different quantities from top to bottom: density (in $M_{\odot}\ pc^{-3}$), $V_\phi$,  $\sigma_R$, $\sigma_\phi$, $\sigma_z$, $\sigma_R/\sigma_z$ (in \kms), tilt angle of the ellipsoid (in degree). Isocontours are spaced by decimal logarithm of the density for densities larger than $10^{-20}\ M_{\odot}\ pc^{-3}$.}
\label{fig:Rz-dens}
\end{center}
\end{figure}
\end{landscape}

Figure \ref{fig:Rz-dens} shows how the density and kinematical parameters (mean rotation velocity, velocity dispersions, and tilt of the ellipsoid) vary for the disc components as a function of Galactocentric coordinates in the final model with declining rotation curve. Similar curves are available upon request for the model with the flat rotation curve. The values are reliable for $|z|<5 kpc$. For the young populations with shorter scale heights, the computation has been performed only to 3 kpc in $z$.

We point out a specific feature seen at high $z$ and $R>15$ kpc in $\sigma_\phi$ in Fig~\ref{fig:Rz-dens} in third row.
The  distribution functions of density and velocities are reliable below $z< 5$ or 6 kpc when $V_{\phi}$ is larger than 120 \kms\ (i.e. the asymmetric drift is smaller than about 110 \kms\ ).  At larger $z$ values the distribution functions are numerically exact and stationary but they are not realistic. This is a consequence of the properties of the generalised Shu distribution function that depends only on positive values of the angular momentum. Then, negative values of the azimuthal Galactic velocity $V_\phi$  or counter rotating stars are not modelled.  When the asymmetric drift  is large, i.e. $V_{\phi}$ smaller than about 120 \kms, corresponding approximately to $z$ larger than 5-6 kpc depending on the population, then the bell shape of the $V_\phi$ distribution is distorted with a maximum closer and closer to $V_{\phi}=0$ and no negative values. A consequence is the `beak' feature seen in Figure 
\ref{fig:Rz-dens}: when $z$ increases, first $\sigma_{\phi}$ increases and beyond $z \sim 6$ kpc it decreases because the $V_\phi$ distribution is `squeezed' towards $V_\phi=0$ values. This problem also impacts the distribution of $\sigma_R/\sigma_z$ in Fig~\ref{fig:Rz-dens} in penultimate row at $z>5$ kpc and $R>$15 kpc. 

\newpage

\section{Comparisons between model and Gaia data selection}
\label{App:comp}

\begin{figure*}[h!]
 \begin{center}
\includegraphics[width=9cm]{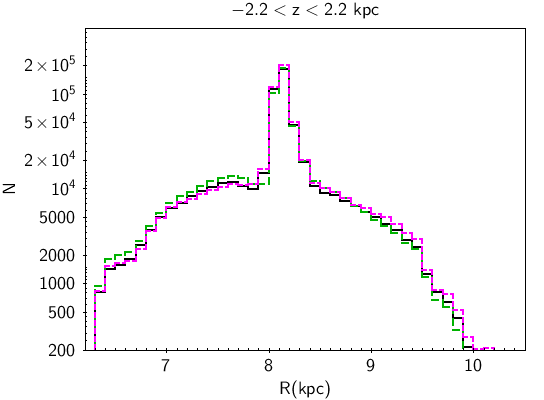}
\includegraphics[width=9cm]{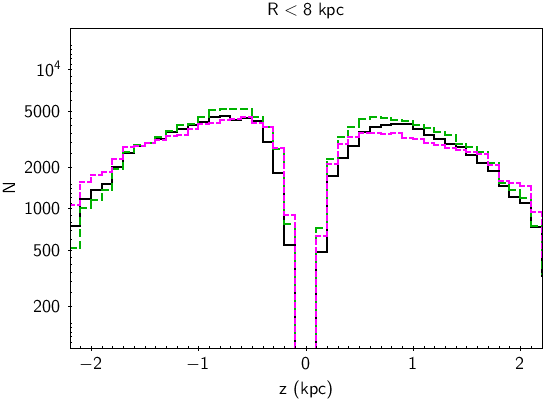}

\includegraphics[width=9cm]{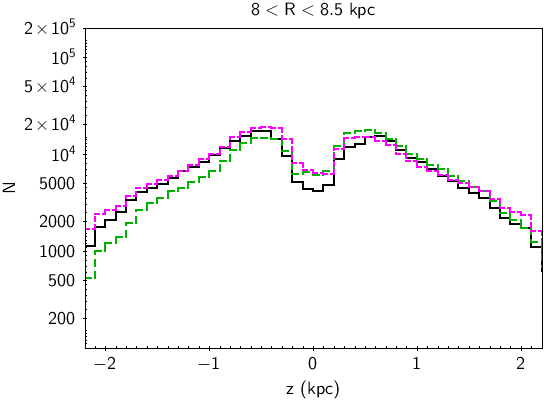}
\includegraphics[width=9cm]{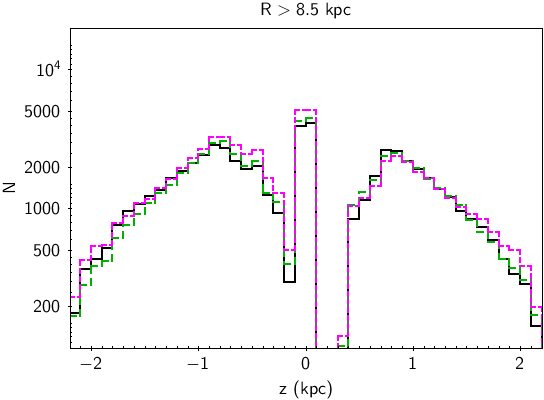}
\caption{Histograms of star counts in the selected sample as a function of pseudo $R$ (top left), pseudo $z$ in inner Galaxy ($R<$8 kpc, top right), at $8<R<$8.5 kpc excluding the local sample (bottom left), towards outer Galaxy ($R>$8.5, bottom right).
\Gaia\  data (black), previous BGM (green long dash), fitted dynamical model (magenta short dash).
}
\label{fig:hist-Rz-comp}
\end{center}
\end{figure*}

In order to ascertain the quality of the fit, we compared stellar densities in the data sample with simulations from model Mev2011 and those from our final fitted dynamical model (Fig. \ref{fig:hist-Rz-comp}). The new model provides much better fit to the radial and vertical distribution seen in the data sample. The total likelihood of the new model is $-902$ (for the declining rotation curve), and $-955$ (for the flat rotation curve), while the one of the Mev2011 model was $-2057$. It remains some significant disagreements in a few areas of the $(R,z)$ plane. The model with its assumptions of axisymmetry however finds  a reasonable mitigation between regions where it overestimates and those where it underestimates the density. Moreover, the model does not account for substructures, like for example the Gaia-Enceladus-Sausage \citep{2018Natur.563...85H,2018MNRAS.478..611B}, or the substructures in the anticentre \citep{2021A&A...646A..99R}. In particular the excess of stars in the model at $R>8500$ pc is solely due to stars at longitude 180\degr\, and $-20\degr<b<20$\degr. It is most probably due to the substructures already known in this region, such as TriAnd, ACS or Monoceros overdensities. These figures anyhow show that the new self-consistent model provides a significant improvement over the previous model Mev2011.

\begin{figure*}[h!]
\begin{center}
\includegraphics[width=9cm]{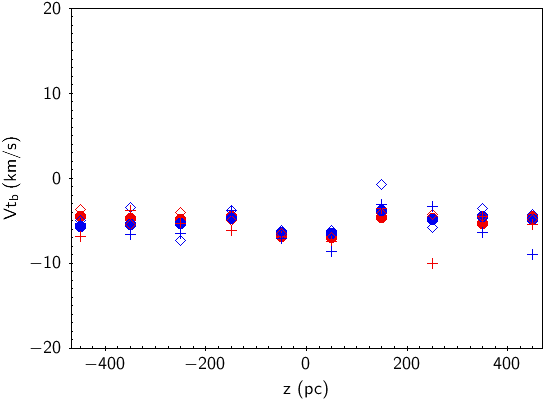}
\includegraphics[width=9cm]{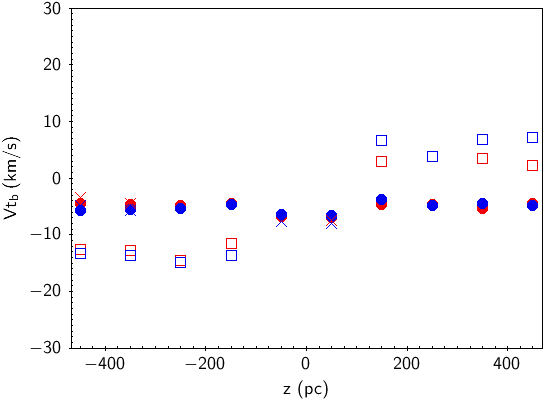}

\includegraphics[width=9cm]{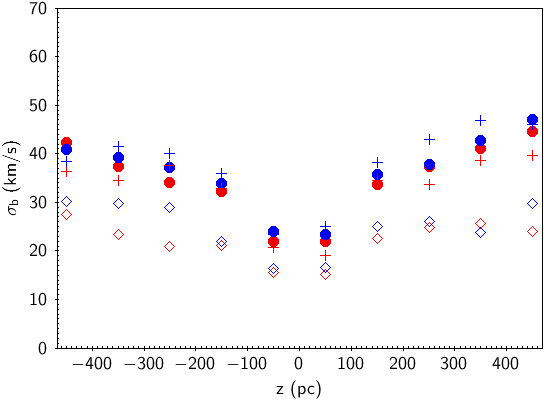}
\includegraphics[width=9cm]{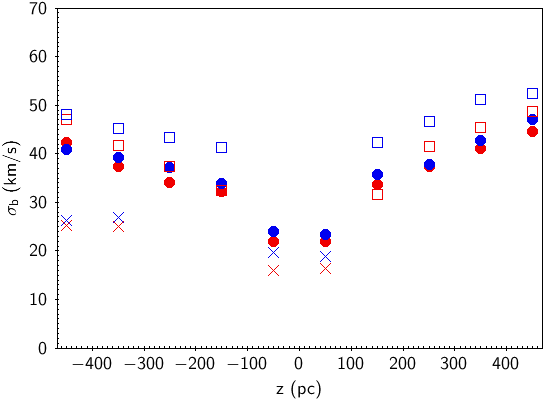}
\caption{Median (top row) and dispersion (bottom row)  of the transverse velocity \VTb\  as a function of $z$ for different subsamples of the final model (red symbols) and data (blue symbols). Full circles are for the whole sample. 
Left panels: Early type stars (\MG$<4$ and \GRP$<0.55$) (open diamonds), giants  (\MG$<$4 and \GRP$>=0.55$) (crosses). {Right panels}:  Selection made on  Galactocentric radii with $R<8$ kpc (open square) and  $R>9$ kpc (x). Points where the number of stars is smaller than ten are considered as insignificant and are not presented.}
\label{fig:sigVTb}
\end{center}
\end{figure*}

\begin{figure*}[h!]
\begin{center}
\includegraphics[width=9cm]{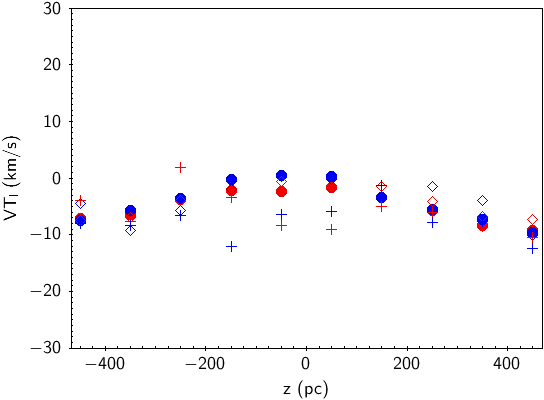}
\includegraphics[width=9cm]{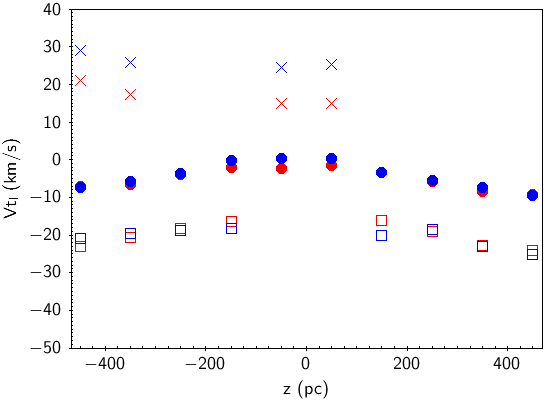}

\includegraphics[width=9cm]{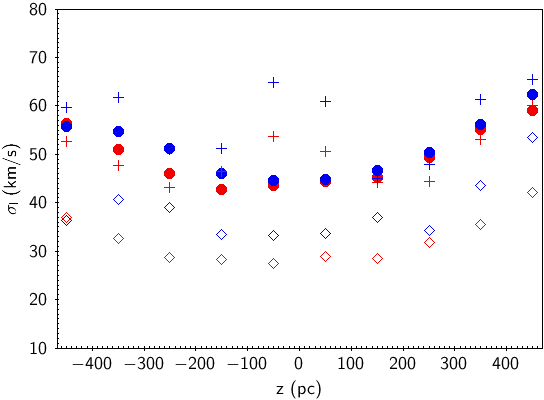}
\includegraphics[width=9cm]{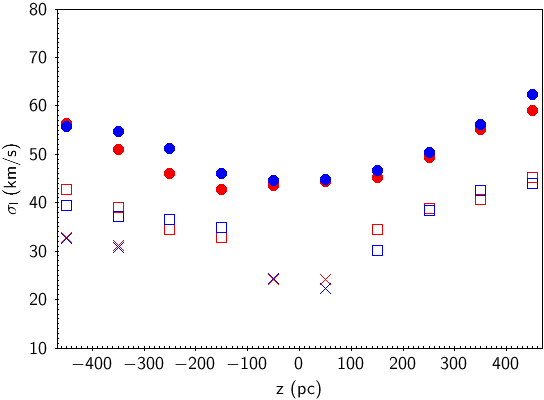}
\caption{
As in Figure~\ref{fig:sigVTb} but for the transverse velocity component \VTl.
}
\label{fig:sigVTl}
\end{center}
\end{figure*}

Medians and standard deviation of the tangential velocities, for the data sample (blues symbols) and the model (red symbols) are presented in Fig.\ \ref{fig:sigVTb} and \ref{fig:sigVTl} for the \VTb\ and \VTl\ component respectively.
In Figure \ref{fig:sigVTb}, we compare the median velocities and velocity dispersions of the transverse velocity \VTb \, as a function of $z$ distance from the Galactic plane between model and data. There is an overall good agreement between model and data. When we split the data  by stellar types (left panel), samples selected by pseudo-absolute \MG\ magnitude and colours allow us to evaluate the change in dispersion for various types according to observational quantities only. It shows that the velocity dispersions for giants roughly (selected by \GRP$>0.55$ and \MG$<4$) vary as a function of \zgal\ as expected from the model, with some noisy fluctuations due to the sample size and local fluctuations (visible when we compare the north and the south for example, also in \cite{2020A&A...643A..75S}), which might be due to non-axisymmetric structures. The selection of early type stars (with \GRP$<0.55$ and \MG$<4$) shows a comparable trend in the model and data, validating the age-velocity dispersion. Right panel demonstrates the effect of change in the velocity dispersion according to Galactocentric radius \Rgal. The model reproduces the overall trend in \zgal\  very well. 

At \Rgal$<8$ kpc the agreement is good, validating the mean dispersion and its radial gradients. The sample at \Rgal$>9$ kpc exhibits smaller velocity dispersions both in model and data  (x symbols in right panels) compared to the inner Galaxy (squares) where the model slightly underestimates the dispersion in $\sigma_b$, although globally (full circles) the agreement between model and data is good with deviations at the level of a few \kms. 

Comparisons of the medians and dispersions of \VTl\ are shown in Figure~\ref{fig:sigVTl}. As for \VTb\ the dispersions from the model are very similar to the data, showing the increase of the dispersion with $z$ and a decrease of the mean velocity due to the increase of the asymmetric drift with distance from the plane. Moreover the medians of \VTl\ are very well reproduced at different Galactocentric radii, validating our radial scale lengths and the asymmetric drift modelled. However, we note that the dispersion is slightly too low in the model (but only by less than 10\%) out of the plane, which is a result of the compromise between fitting density and kinematics.

\end{appendix}


\end{document}